\documentclass[journal,twocolumn]{IEEEtran}
\pdfoutput=1
\newlength{\figwidth}
\usepackage{color}
\definecolor{links}{rgb}{0.7,0,0}   
\definecolor{urls}{rgb}{0,0,0.8}    
\definecolor{cites}{rgb}{0,0,0.8}   

\usepackage[colorlinks,hyperindex,linkcolor=links,citecolor=cites,urlcolor=urls]{hyperref} 

\usepackage[nosort]{cite}
\usepackage{url}
\usepackage[intlimits]{amsmath}
\usepackage{bbm}
\usepackage{graphicx}
\usepackage{paralist}
\usepackage{fancyref}
\usepackage[stretch=16,shrink=16,step=4]{microtype}
\usepackage{vmr-symbols-vecbold}
\usepackage{standard-macros}
\usepackage{siunitx}
\usepackage[font=footnotesize]{caption} 
\usepackage[font=footnotesize]{subcaption}
\usepackage{glossaries}
\glsdisablehyper
\newacronym{rf}{RF}{radio frequency}
\newacronym{cpu}{CPU}{central processing unit}
\newacronym{mimo}{MIMO}{multiple-input multiple-output}
\newacronym{ap}{AP}{access point}
\newacronym{adc}{ADC}{analog-to-digital converter} 
\newacronym{dsp}{DSP}{digital signal processing} 
\newacronym{bb}{BB}{base-band}
\newacronym{bp}{BP}{band-pass} 
\newacronym{osr}{OSR}{oversampling rate}
\newacronym{ue}{UE}{user equipment} 
\newacronym{ofdm}{OFDM}{orthogonal frequency-division multiplexing} 
\newacronym{lmmse}{LMMSE}{linear minimum mean-squared error} 
\newacronym{evm}{EVM}{error vector magnitude} 
\newacronym{mr}{MR}{maximum ratio} 
\newacronym{zf}{ZF}{zero forcing} 
\newacronym{bs}{BS}{base station} 
\newacronym{csi}{CSI}{channel state information} 

\setacronymstyle{long-short}

\makeatletter
\def\@IEEEinterspaceratioM{0.265}
\def\@IEEEinterspaceMINratioM{0.1651}
\def\@IEEEinterspaceMAXratioM{0.38}

\def\@IEEEinterspaceratioB{0.31}
\def\@IEEEinterspaceMINratioB{0.19}
\def\@IEEEinterspaceMAXratioB{0.38}
\@IEEEtunefonts
\makeatother
\begin{document}

\IEEEoverridecommandlockouts

\title{EVM Analysis of Distributed Massive MIMO with 1-Bit Radio-Over-Fiber
	Fronthaul}
%
%

\author{Anzhong Hu, \IEEEmembership{Member, IEEE}, Lise Aabel, Giuseppe Durisi,
	\IEEEmembership{Senior Member, IEEE},  Sven Jacobsson, Mikael
	Coldrey, Christian Fager,~\IEEEmembership{Senior~Member,~IEEE}, Christoph Studer,~\IEEEmembership{Senior~Member,~IEEE}
	\thanks{The work of L. Aabel, C. Fager, and G. Durisi was supported in part by the Swedish Foundation for Strategic
		Research under Grant ID19-0036. This research has also been carried out in part within the
		Gigahertz-ChaseOn Bridge Center, in a project financed by Chalmers, Ericsson, and Qamcom.
	}

	\thanks{Anzhong Hu is with the
		School of Communication Engineering, Hangzhou Dianzi University, Hangzhou, China (e-mail:
		huaz@hdu.edu.cn).}
	\thanks{Lise Aabel, Sven Jacobsson and Mikael Coldrey are with Ericsson AB, Gothenburg, Sweden.
		(e-mail: \{lise.aabel,sven.jacobsson, mikael.coldrey\}@ericsson.com).}
	\thanks{Giuseppe Durisi and Christian Fager are with Chalmers University of Technology,
		Gothenburg, Sweden (e-mail: \{durisi,christian.fager\}@chalmers.se).}
	\thanks{Christoph Studer is with the Department of Information Technology and Electrical Engineering, ETH Zurich, Zurich, Switzerland (e-mail: studer@ethz.ch).}
	\thanks{This paper was presented in part at the IEEE Global Telecommunication Conference (GLOBECOM), Waikoloa, HI, USA, 2019~\cite{jacobsson19-07a}.
	}
}

\maketitle


\begin{abstract}
	We analyze the uplink performance of a distributed massive multiple-input
	multiple-output (MIMO) architecture in which the remotely located access
	points (APs) are connected to a central processing unit via a fiber-optical
	fronthaul carrying a dithered and 1-bit quantized version of the
	received radio-frequency (RF) signal. The innovative feature of the proposed
	architecture is that no down-conversion is performed at the APs. This
	eliminates the need to equip the APs with local oscillators, which may be
	difficult to synchronize. Under the assumption that a constraint is imposed on
	the amount of data that can be exchanged across the fiber-optical fronthaul,
	we investigate the tradeoff between spatial oversampling, defined in
	terms of the total number of APs, and temporal oversampling,
	defined in terms of the oversampling factor selected at the central processing
	unit, to facilitate the recovery of the transmitted signal from
	1-bit samples of the RF received signal. Using the so-called error-vector
	magnitude (EVM)  as performance metric, we shed light on the optimal design
	of the dither signal, and quantify, for a given number of
	APs, the minimum fronthaul rate
	required for our proposed distributed massive MIMO
	architecture to outperform a standard co-located massive MIMO
	architecture in terms of EVM.
\end{abstract}
%

\section{Introduction}\label{sec:introduction}
Distributed massive \gls{mimo} (also referred in the literature as cell-free
massive \gls{mimo}), refers to an architecture, in which a large number of
\glspl{ap} serve, in a coordinated way, a much smaller number of \glspl{ue} over a
given coverage area~\cite{ngo17-03a,nayebi17-07x,demir20-a}. Coordination is
provided by a \gls{cpu} connected to the \glspl{ap} via fronthaul links. The
most effective form of coordination is achieved when the \gls{cpu} performs all
digital processing tasks (including channel estimation and spatial processing)
up to digital-to-analog and analog-to-digital conversion, whereas the APs
perform only analog processing, including up- and
down-conversion~\cite{bjornson20-01a}.

Distributed massive \gls{mimo} architectures have been shown to be advantageous
over more traditional co-located massive \gls{mimo} architectures, where
a single \gls{bs}, equipped with a large number of
co-located antennas, serves all users in the coverage area. Indeed, when equipped with
\gls{lmmse} spatial filtering and fully-centralized processing, distributed massive
MIMO solutions are able to exploit macro-diversity, and can mitigate the path-loss
variations experienced in co-located massive \gls{mimo} architectures. This
results in a more uniform quality of service across the coverage area compared to
co-located solutions~\cite{bjornson20-01a}.
To make distributed massive MIMO practical, serial network topologies and distributed algorithms to reduce
fronthaul signaling have been recently investigated~\cite{rodriguez-sanchez20-01a,shaik21-11a}.

\subsubsection*{Practical Challenges}\label{sec:practical-challenges}
A crucial and often implicit assumption behind the theoretical analyses
illustrating the advantages of distributed massive \gls{mimo} architectures over
co-located architectures is that the \glspl{ap} are able to operate in a phase-coherent
way when performing up- and down-conversion of the \gls{rf} signals. In
practice, however, each \gls{ap} will need to be equipped with an individual
oscillator. This will result in independent phase noise during up- and
down-conversion, which, if not compensated for, will affect performance significantly.
Synchronizing such oscillators via the distribution of a common clock is
challenging even in co-located massive \gls{mimo} architectures, especially at
millimeter-wave frequencies~\cite{rasekh21-10a}. Such an approach appears even
more challenging in
distributed deployments, also at lower frequencies. Hence, transmission
resources, in the form of control signals for synchronization purposes, need
to be sacrificed to make the system operate coherently.
Performing such a synchronization over a large number of \glspl{ap} is, however, a nontrivial task
and guaranteeing a low synchronization error may not even be possible in certain distributed
\gls{mimo}
deployments~\cite{larsson24-01a}.

This issue has recently motivated the exploration of an alternative hardware
architecture, in which \gls{cpu} and \glspl{ap} are connected via a fiber-optical
fronthaul, and the independent phase-noise resulting from the use of individual
oscillators at the \glspl{ap} is avoided by letting the \gls{cpu}, instead of
the \gls{ap}, perform
up- and down-conversion~\cite{aabel20-11p,sezgin21-02a}. Since transferring
analog \gls{rf} signals over a fiber-optical fronthaul is challenging
due to the high linearity requirements, the authors
of~\cite{aabel20-11p,sezgin21-02a}, building on results previously reported
in~\cite{cordeiro17-11c,wang19-06u,wu20-02a,wu20-05b}, propose to map a dithered
version of the \gls{rf} signal into a
two-level square waveform, before transmission
over the fronthaul link. Specifically, in the downlink of the
architecture proposed and experimentally demonstrated via a testbed in~\cite{aabel20-11p,sezgin21-02a}, the \gls{cpu}
generates the desired \gls{bb} digital signals and up-converts them
into temporally oversampled 1-bit digital \gls{rf} signals by means of a
sigma-delta modulator. The resulting two-level analog signals at the output of
the
digital-to-analog converters are then transferred
to the single-antenna \glspl{ap} via the fiber-optical fronthaul, where
they are
\gls{bp} filtered (to remove quantization noise), amplified, and fed to the
antenna element.

In the uplink, the received signal at each \gls{ap} is amplified, filtered, and
then converted into a two-level analog signal by means of a comparator, whose
second port is connected to a dither signal generated at the
\gls{cpu} and conveyed to the \glspl{ap} via the downlink fiber-optical
fronthaul.
The resulting signal is sent to the \gls{cpu} via the uplink
fiber-optical fronthaul, where it is sampled and fed to the \gls{dsp} unit.
The testbed measurements reported
in~\cite{aabel20-11p,sezgin21-02a,aabel23-12a} show that this architecture, which can be
implemented using low-cost, off-the-shelf components, and which we shall refer to in the
remainder of the paper as \emph{distributed massive \gls{mimo} with 1-bit
	radio-over-fiber fronthaul}, yields satisfactory performance in terms of
\gls{evm}.

The purpose of this paper is twofold: we present a theoretical framework for the
analysis of the uplink \gls{evm} attainable with the distributed massive \gls{mimo}
with 1-bit radio-over-fiber fronthaul architecture presented
in~\cite{aabel20-11p,sezgin21-02a,aabel23-12a}, and use this
framework to provide insights into the optimal choice of system parameters such as
the power of the dither signal,
the required fronthaul capacity, the number of \glspl{ap}, and the
\gls{osr} to be employed at the \gls{cpu} to reconstruct the \gls{rf} signal from the two-level waveform
transmitted over the fiber-optical fronthaul.

\subsubsection*{State of the Art}
The problem of analyzing the performance of co-located massive \gls{mimo}
architectures in the presence of nonlinearities caused by low-precision
converters is well-studied in the literature (see,
e.g.,~\cite{jedda16a,mollen17-01a,jacobsson17-06a,li17-08a,jacobsson19-03b}). In
this line of work, low-precision quantizers are introduced to limit the power
consumption and to alleviate the throughput requirements on the fronthaul links connecting
the (single) radio unit to the \gls{cpu}. The main finding in these papers is that
satisfactory performance can be achieved both in the uplink and in the downlink,
provided that the number of antennas at the \gls{bs} is sufficiently large, and
that dithering is used to whiten the quantization noise, whenever the signal-to-interference-and-noise ratio
is large.

Some of these findings have recently been extended to the distributed massive
\gls{mimo} setting.
Specifically, low-precision quantizers are introduced in this setting to model a constraint
on the capacity of the fronthaul links connecting the \glspl{ap} to the
\gls{cpu},
and similar conclusions, in terms of achievable performance, are
reached~\cite{zhang19-a,hu19-10a,bashar19-12a,femenias20-a,bashar21-09a}.
However, in these works, a \emph{homodyne} transceiver architecture is assumed, which
means that quantization is
performed on the real and imaginary parts of the complex-valued \gls{bb} signal
(either before or after spatial processing at the \gls{ap}),
whereas in the present paper we will consider an architecture in which quantization
is performed directly on the real-valued \gls{rf} (i.e., \gls{bp}) signal and spatial processing
is performed entirely at the \gls{cpu}.

To achieve satisfactory performance with the architecture proposed
in~\cite{aabel20-11p,sezgin21-02a}, it is imperative that the two-level signal carrying a
1-bit quantized version of the \gls{rf} signal is sampled
at a much higher rate than the one required for the \gls{bb}
signal. The benefits of oversampling in the presence of $1$-bit converters have
been pointed out previously in the literature (see, e.g.,~\cite{landau18-10p,schluter20-10e}),
although only for scenarios in which quantization is
performed on the \gls{bb} signal.
Hence, the question whether spatial oversampling obtained via the deployment of
multiple \glspl{ap} can mitigate the need for temporal oversampling remains largely
unexplored in the literature.
\subsubsection*{Contributions}
We consider a distributed multiuser massive \gls{mimo} system in
which~$U$ \glspl{ue} are served by $B$ distributed \glspl{ap} connected to a
\gls{cpu} via a 1-bit radio-over-fiber fronthaul,
and characterize the performance, in terms of \gls{evm}, achievable in the uplink.
Our characterization relies on an application of Bussgang's
decomposition~\cite{bussgang52a} on a frequency-domain version of the \gls{rf} signal received at each
\gls{ap}.
This allows us to express the oversampled 1-bit-quantized
discrete-time \gls{rf} signal at the \gls{cpu} as a linear function of the
transmitted \gls{bb} signal.

For a given fronthaul-rate constraint, which models the processing speed of the
\glspl{adc} at the \gls{cpu},
and for the case in which the dither signal is modeled as a Gaussian process,
we provide expressions for the \gls{evm} achievable using different linear
spatial combiners developed on the basis of the Bussgang decomposition,
including \gls{mr}, \gls{zf}, and (quantization-aware)
\gls{lmmse} combiners.
We also provide an analytical characterization of the \gls{evm} in the asymptotic
limit of large fronthaul rate, and interpret it as the \gls{evm} of an
equivalent distributed \gls{mimo} homodyne architecture.
This analytical characterization turns out to provide relevant insights into the
optimal design of the dither signal.

Focusing on a scenario in which the \glspl{ue} transmitted power is
inversely proportional to the number of \glspl{ap}, to avoid favoring unfairly deployments
with high \gls{ap} density
and encourage energy efficiency,\footnote{More sophisticated models to analyze the
	energy efficiency of distributed \gls{mimo} architectures are available in the
	literature~\cite{bashar19-12a}. However, these models pertain scenarios in which
	part of the signal
	processing is performed at the \glspl{ap}, and, hence, are not applicable to the
	architecture considered in the present paper.}
we conduct numerical
experiments that confirm the correctness of the insights obtained via the
asymptotic analysis, and shed light both on the impact of the fronthaul-rate constraint, and
on the tradeoff between temporal and spatial oversampling.
Here, temporal oversampling is
dictated by the \gls{osr} at the \gls{cpu}, whereas spatial
oversampling is dictated by the number of \glspl{ap} deployed in the coverage area.

Our numerical results suggest that, under the transmitted-power normalization considered
in the paper, when a single \gls{ue} is active, an \gls{mr}/\gls{zf} combiner
is used, and the
fronthaul rate is small,
temporal oversampling is preferable to spatial oversampling.
Specifically, the available fronthaul rate should be used to acquire, from few
\glspl{ap}, a highly
oversampled version of the two-level \gls{rf} signal.
For the case of \gls{lmmse} combiner, however, one can trade spatial oversampling
for temporal oversampling in this regime.
As the fronthaul rate increases, spatial oversampling
becomes preferable for all three combiners.
This is because, in the asymptotic regime of infinitely large fronthaul
rate, the \gls{evm} decreases monotonically with the number of \glspl{ap}.
When multiple users are active, spatial oversampling is more desirable than
temporal oversampling, even when the
fronthaul rate is small, since spatial oversampling reduces multiuser interference.

\subsubsection*{Notation}
Lower-case boldface letters are used for vectors and upper-case boldface letters for matrices.
We denote by  $\jpg(\veczero_{N} ,\matR)$ the distribution of an
$N$-dimensional circularly symmetric complex-valued Gaussian
vector with zero mean and  $N\times N$ covariance matrix $\matR$. We use $\matI_{N}$ to denote the $N\times N$ identity
matrix, $\veczero_{N}$ to denote the $N\times 1$ zero vector, $\veczero_{N,N}$ to denote $N\times N$ zero matrix, and $\Ex{}{\cdot}$ and $\Prob\lefto[\cdot\right]$ to denote expectation
and probability, respectively.
For a sequence of random variables $\{x_{n}\}$, we use $x_{n}
	\overset{\text{a.s.}}{\longrightarrow} x$, $n\to\infty$, where a.s. stands for
almost surely, to indicate that
$\Prob\lefto[\lim_{n\to\infty} x_{n}=x  \right]=1$.
The superscripts $\tp{}$, $\conj{}$,  and $\herm{}$ stand for transposition,
element-wise conjugation, and Hermitian transposition, respectively.
The
Kronecker delta function $\delta[n]$ is defined as $\delta[0]=1$ and $\delta[n]=0$ for $n\neq
	0$, $\floor{x}$ is the largest integer that is smaller or equal to
$x\in\reals$, and $\mathrm{sgn}(x)$ denotes the sign function, which is equal to
$-1$ if $x<0$ and to $1$ if $x\geq 0$.
For a matrix $\matA$, we let $\tr(\matA)$ denote its trace and $\diag(\matA)$
a diagonal matrix whose diagonal entries coincide with the diagonal
entries of $\matA$.
Finally, we use $[\matA]_{b,b'}$ to indicate the entry on row~$b$ and column $b'$
of $\matA$.
\section{System Model}
\label{sec:system-model}
\begin{figure}[!t]
	\centerline{{\includegraphics[width=1.0\figwidth]{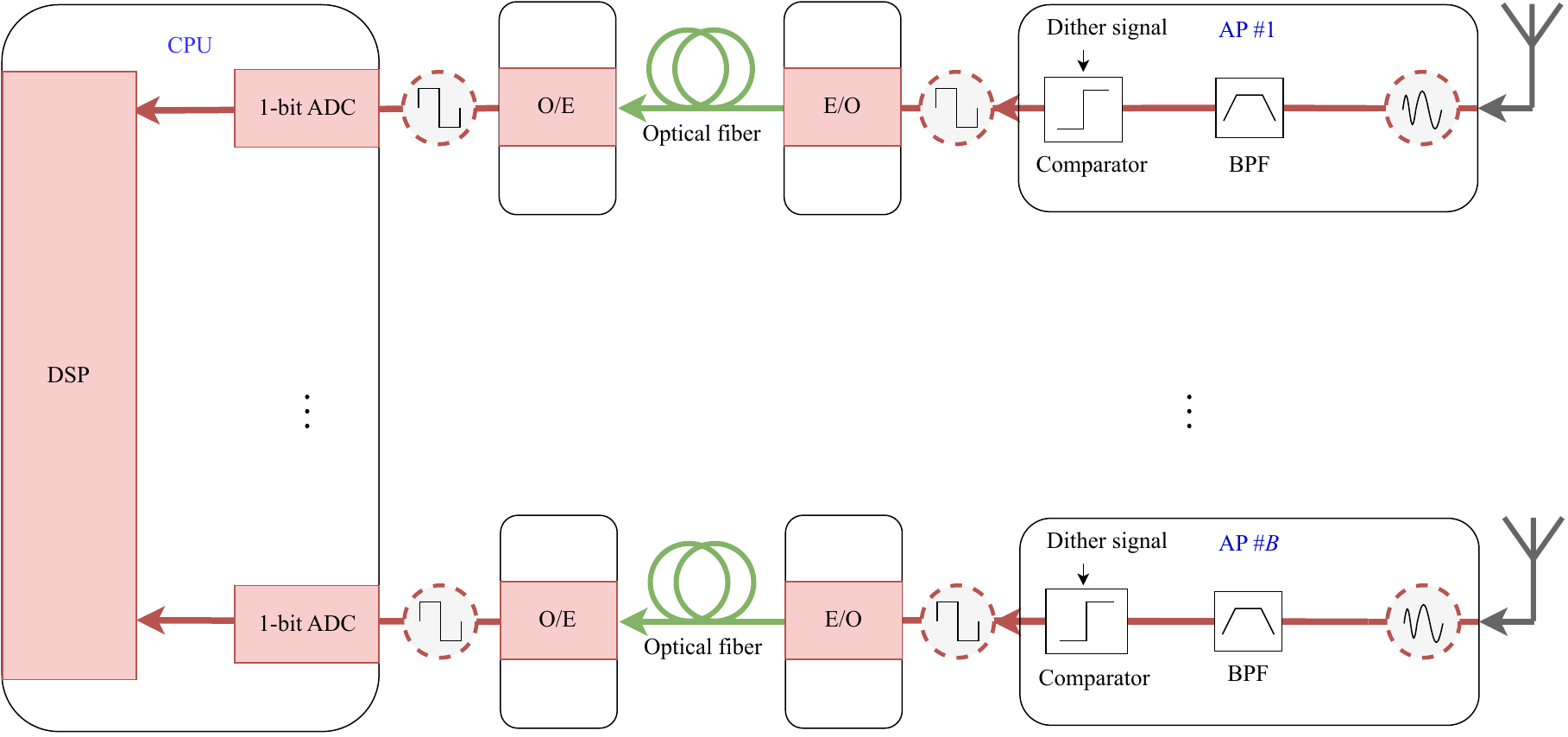}}}
	\caption{A distributed massive \gls{mimo} uplink architecture consisting
		of $B$ \glspl{ap} connected to a
		\gls{cpu} via a fiber-optical fronthaul.
		The received signal at each AP is \gls{bp} filtered, added to a
		dither
		signal and compared with a zero
		threshold.
		The resulting two-level signal is converted to the optical domain and transmitted over the optical
		fiber to the \gls{cpu}, where it is converted back to the electrical domain,
		oversampled using a $1$-bit \gls{adc}, and sent to the \gls{dsp} for digital down-conversion and spatial
		processing.
	}
	\label{fig:system}
\end{figure}
The distributed \gls{mimo} architecture considered in this paper and depicted in
Fig.~\ref{fig:system}, follows the structure
proposed in~\cite{aabel20-11p,sezgin21-02a}.
Specifically, the \gls{rf} signal received at the single-antenna port of $B$
\glspl{ap} is \gls{bp} filtered (BPF) and added to a dither signal.
The resulting signal is then passed through a zero-threshold comparator, which
produces a two-level waveform.
This two-level waveform is converted into the optical domain by an electrical-to-optical (E/O) converter and conveyed to the
\gls{cpu} via a
fiber-optical fronthaul.
At the \gls{cpu}, the received signal from each \gls{ap} is converted back to the electrical domain by an
optical-to-electrical (O/E) converter,
oversampled by a $1$-bit \gls{adc}, and sent to the \gls{dsp} for digital down-conversion, spatial processing,
and demodulation/decoding.
\subsection{Signal Parameters and Assumptions}\label{sec:signal-system}
We consider the transmission of a \gls{bp} signal of bandwidth $W$ centered at
frequency $f\sub{c}\gg W$.\footnote{For convenience, we shall assume throughout
	that the ratio $f\sub{c}/W$ is an integer.}
The corresponding received signal at each \gls{ap} is passed through an ideal
\gls{bp} filter with bandwidth $W$ centered at $f\sub{c}$ to remove out-of-band noise and
interference.
We denote by $f\sub{s}$ the sampling frequency of the \glspl{adc} at the
\gls{cpu} and focus on the scenario in which $f\sub{s}\gg W$.
Indeed, since we do not perform down-conversion in the analog domain, but only
digitally at the \gls{cpu}, sampling
the received signal at the Nyquist
rate $2W$ for the underlying received \gls{bb} signal may not result in perfect
reconstruction, even when no quantization is performed.
We will also not necessarily require that $f\sub{s}\geq 2f\sub{c}+W$,
since it is possible to reconstruct (again, for the case of no
quantization) the underlying \gls{bb} received signal from samples of the
\gls{bp} signal taken at any rate $f\sub{s}$ satisfying~\cite{vaughan91-09a}
\begin{equation}\label{eq:nyquist-bandpass}
	(2f\sub{c}+W)/\ell \leq f\sub{s} \leq (2f\sub{c}-W)/(\ell-1),
\end{equation}
where the integer $\ell$ belongs to the set $\{ 1,2,\dots,
	\floor{(f\sub{c}+W/2)/W}\}$.
Note that the choice $\ell=1$ results in $f\sub{s}\geq 2f\sub{c}+W$.

Departing from the solution implemented in the testbed described in~\cite{aabel20-11p,sezgin21-02a}, we shall model, for
analytical convenience, the dither signal as a white Gaussian noise process of double-sided spectral density $E\sub{d}/2$ with
respect to the bandwidth $f\sub{s}$ (see~\cite[Def. 25.25.1]{lapidoth09a}).\footnote{In the testbed described
	in~\cite{aabel20-11p,sezgin21-02a,aabel23-12a}, the dither signal is a $\qty{17}{MHz}$ triangular waveform, generated at the \gls{cpu} via
	sigma-delta modulation, and conveyed
	to the \glspl{ap} via the downlink fronthaul link.
	As shown in~\cite{jacobsson19-08a}, a Gaussian dither signal, which is more difficult to implement,
	results in slightly better \gls{evm}. So our analysis may slightly under-estimate the \gls{evm} that
	would be achievable in practice.}

\subsection{Discretized Input-Output Relation}
For a fixed $N$, we focus on the signal received over a time interval of length $T=N/f\sub{s}$, and write
the $n$th sample at the output of the $B$ $1$-bit \glspl{adc}  at the \gls{cpu} as
\begin{equation}\label{eq:rf-quantized}
	\vecz_{n}\supp{RF} = \mathrm{sgn}\bigl(\vecy_{n}\supp{RF} + \vecd_{n}\bigr), \quad
	n=0,\dots,N-1.
\end{equation}
Here, $\vecy_{n}\supp{RF} \in \reals^{B}$ denotes the $n$th sample of the discrete-time \gls{rf} signal before $1$-bit
quantization but after \gls{bp} filtering and sampling, $\vecd_{n}\distas \normal\lefto(\veczero_B, (E\sub{d}/2)\matI_{B}\right)$
is the Gaussian dither signal (after sampling), which we assume to be independent of $\vecy_{n}\supp{RF}$ and independent across $n$, and the $\mathrm{sgn}(\cdot)$ function is applied element-wise to its vector-valued input.
We model $\vecy_{n}\supp{RF}$ as
\begin{equation}\label{eq:RF-nonquantized-output}
	\vecy_{n}\supp{RF} = \sqrt{2} \Re\bigl\{ \vecy_{n}\supp{BB} e^{j2\pi (f\sub{c}/f\sub{s}) n}\bigr\},
\end{equation}
where $\vecy_{n}\supp{BB} \in \complexset^{B}$ denotes the complex envelope of the discrete-time received signal.

Let now $S=WT$, and assume for simplicity that $S$ is an odd integer.
Define the following set:
\begin{multline}\label{eq:setS}
	\setS = \{0,1,\dots, (S-1)/2, N-(S-1)/2, N-(S-1)/2+1,\\ \dots, N-1 \}.
\end{multline}
To take into account the correlation introduced on the \gls{bb} signal
$\vecy_{n}\supp{BB}$ by oversampling, we shall model it via an inverse discrete Fourier transform
as\footnote{For simplicity, we shall not consider phase noise or other forms of impairments at the \glspl{ue}.}
\begin{equation}\label{eq:received-signal-frequency-domain}
	\vecy_{n}\supp{BB} = \frac{1}{\sqrt{N}} \sum_{k\in \setS} \bigl( \widehat{\matH}_{k} \hat{\vecs}_{k} + \hat{\vecw}_{k}
	\bigr) e^{j2\pi \frac{k}{N}n},
\end{equation}
where $\hat{\vecs}_{k} \in \complexset^{U}$  denotes the discrete signal, expressed in the frequency domain,
transmitted by the $U$ \glspl{ue}, $\hat{\vecw}_{k}\distas \jpg(\veczero_B,
	N_{0} \matI_{B})$, $k\in\setS$, denotes the
band-limited additive Gaussian noise at the \glspl{ap}, and $\widehat{\matH}_{k}\in
	\complexset^{B\times U}$ denotes the channel frequency
response.
Throughout, we shall assume that $\hat{\vecs}_{k}\distas \jpg(\veczero_U, E\sub{s} \matI_{U})$, where $E\sub{s}$ denotes
the average energy per sample, and that the vectors $\{\hat{\vecs}_{k}\}_{k\in\setS}$ and $\{\hat{\vecw}_{k}\}_{k\in \setS}$ are
independent, whereas we will allow for arbitrary dependence among the matrices $\{\widehat{\matH}_{k}\}_{k\in\setS}$.
Throughout most of the paper, we shall assume for simplicity that the matrices $\{\widehat{\matH}_{k}\}_{k\in\setS}$ are perfectly known
to the \gls{cpu}.\footnote{The case of imperfect channel knowledge will
	be discussed in Section~\ref{sec:imperfect-csi}.}
The ratio $O=N/S=f\sub{s}/W$ is the temporal \gls{osr}.

\subsection{Assumptions}
We are interested in how the performance of the distributed multiuser
massive \gls{mimo} architecture with 1-bit radio-over-fiber fronthaul depicted
in Fig.~\ref{fig:system} is influenced by the number
of deployed \glspl{ap}, for a given total coverage area and a given fronthaul
rate.
To perform a fair comparison between architectures with a different number of \glspl{ap}, we
introduce the following two assumptions.

\paragraph*{Fronthaul Constraint}
We assume that there exists a constraint on the number of samples per second
that can be processed at the \gls{cpu}. This imposes a constraint on the total
fronthaul rate of the architecture, which we shall denote by $R\sub{fh}$.
Mathematically, $B  f\sub{s} \leq R\sub{fh}$.
%
The following equivalent formulation of the fronthaul constraint will turn out more insightful.
For a fixed bandwidth $W$ of the transmitted signal, we have that
\begin{equation}\label{eq:fronthaul-constraint}
	B  O \leq R\sub{fh}/W.
\end{equation}
In words, according to~\eqref{eq:fronthaul-constraint}, if we want to deploy more \glspl{ap}, we need to reduce the rate at which
the received \gls{rf} signal at each \gls{ap} is oversampled.

\paragraph*{Energy-Efficient Setup}
For a fixed coverage area, increasing the number of \glspl{ap} implies reducing the distance between each \gls{ue} and the
closest \gls{ap}.
To benefit from this proximity, without favoring unfairly architectures with a large \gls{ap}
density,  and to encourage energy efficiency, we let the transmit power
at each \gls{ue} to be inversely proportional to the number of \glspl{ap} $B$.
Specifically, 
\begin{equation}\label{eq:energy-efficient-setup}
	E\sub{s} = \bar{E}\sub{s}/B
\end{equation}
for a fixed normalized energy per sample $\bar{E}\sub{s}$.
Since we decided to benefit from the \glspl{ap} proximity by reducing the transmit power, we refer to the scenario that results from
the assumption~\eqref{eq:energy-efficient-setup} as energy-efficient setup.\footnote{Under
	the proposed normalization, and for a uniform two-dimensional \gls{ap} deployment scenario within a
	fixed coverage area, the
	\gls{evm} for the case of infinite fronthaul can be shown to converge to a positive value as
	$B\to\infty$.}

\subsection{Linear Decomposition Using Bussgang's Theorem}
Given our assumptions on $\hat{\vecs}_{k}$, $\hat{\vecw}_{k}$, and on the dither signal $\vecd_{n}$, the input
$\vecq_{n} = \vecy_{n}\supp{RF} + \vecd_{n}$
to the
quantizer in~\eqref{eq:rf-quantized} is a conditionally Gaussian vector
given the channel matrices $\{\widehat{\matH}_{k}\}$.\footnote{It turns out that
	$\vecy_n\supp{RF}$ can be well-approximated by a Gaussian random vector also
	for the case in which
	the transmitted signal is drawn from commonly used constellations, provided that \gls{ofdm} is used and the number of
	subcarriers is sufficiently large~\cite{jacobsson17-12b}.}
As a consequence, we can use Bussgang's decomposition~\cite{bussgang52a} to linearize
the input-output relation after quantization---an approach that is
common in the quantized massive \gls{mimo} literature (see, e.g.,~\cite{mollen17-01a,saxena17-09a,jacobsson17-12b,jacobsson19-03b}).
Specifically,  using Bussgang's decomposition,
we can express the quantized signal $\vecz_n\supp{RF}$ in~\eqref{eq:rf-quantized} as
\begin{equation}\label{eq:bussgang-rf}
	\vecz_n\supp{RF} = \matG \vecq_{n} + \vece_n,
\end{equation}
where $\vece_n \in \reals^{B} $ is the quantization distortion, which is uncorrelated
with $\vecq_{n}$, and $\matG \in \reals^{B\times B}$ is the
Bussgang gain matrix, given by
\begin{equation}\label{eq:bussgang-matrix}
	\matG = \sqrt{\frac{2}{\pi}} \diag\lefto(
	\matR_{\vecq}[0]\right)^{-1/2}.
\end{equation}
Here, $\matR_{\vecq}[m] = \Ex{}{\vecq_{n}
		\tp{(\vecq_{n-m})}}$ denotes the
$B\times B$ autocovariance matrix of $\vecq_{n} $.
It will turn out important to characterize this quantity for arbitrary values of $m$.
Since the dither signal is white and  independent of the received signal, we
conclude that
\begin{equation}\label{eq:matrix-R-q-RF}
	\matR_{\vecq}[m]=\matR_{\vecy\supp{RF}}[m] +
	\frac{E\sub{d}}{2}\matI_{B}\delta[m],
\end{equation}
where $\matR_{\vecy\supp{RF}}[m]=\Ex{}{\vecy\supp{RF}_{n}
		\tp{(\vecy\supp{RF}_{n-m})}}$.
Furthermore, it follows from~\eqref{eq:RF-nonquantized-output},~\eqref{eq:received-signal-frequency-domain}, and from the fact that the
$\{\hat{\vecs}_{k}\}$ and $\{\hat{\vecw}_{k}\}$ are independent and complex proper Gaussian (so that, in particular,
$\Ex{}{\hat{\vecs}_{k}\tp{(\hat{\vecs}_{k})}}=\veczero_{U,U}$ and $\Ex{}{\hat{\vecw}_{k}\tp{(\hat{\vecw}_{k})}}=\veczero_{B,B}$~\cite[p.~502]{lapidoth09a}), that
\begin{multline}\label{eq:autocorrelation-transmitted-signal}
	\matR_{\vecy\supp{RF}}[m]= \\ \frac{1}{N}\Re\lefto\{\sum_{k\in \setS} \bigl(E\sub{s}\widehat{\matH}_k
	\herm{\widehat{\matH}}_k +N_{0}\matI_{B}\bigr) e^{j2\pi (k/N + f\sub{c}/f\sub{s}) m} \right\}.
\end{multline}
Let now $\matR_{\vece}[m] = \Ex{}{\vece_{n} \tp{(\vece_{n-m})}}$ be the autocovariance
matrix of the quantization error and $\matR_{\vecz\supp{RF}}[m] = \Ex{}{\vecz\supp{RF}_{n} \tp{(\vecz\supp{RF}_{n-m})}}$
the autocovariance of the quantized signal in~\eqref{eq:rf-quantized}.
It follows from~\eqref{eq:bussgang-rf} that
\begin{equation}\label{eq:covariance_quantization_error_RF}
	\matR_{\vece}[m] = \matR_{\vecz\supp{RF}}[m] - \matG \matR_{\vecq}[m]\matG,
\end{equation}
where the conditional Gaussianity of $\vecq_n$ implies that $\matR_{\vecz\supp{RF}}[m]$ admits the following closed-form
expression, usually referred to as arcsine-law~\cite[Eq.~(17)]{van-vleck66a}:
\begin{equation}\label{eq:arcsine-law}
	\matR_{\vecz\supp{RF}}[m] = \frac{2}{\pi} \arcsin\lefto(\matD^{-1/2}
	\matR_{\vecq}[m]\matD^{-1/2}\right).
\end{equation}
Here, $\matD = \diag\lefto(\matR_{\vecq}[0]\right)$.

\subsection{Digital Down-Conversion and Linear Combining}
To recover the transmitted signal, the \gls{cpu} down-converts the 1-bit quantized
\gls{rf} signal in the digital domain.
Then, it transforms the resulting BB signal into the frequency domain, and keeps
only the frequency samples in the set $\setS$.
This results in
\begin{equation}
	\hat{\vecz}\supp{BB}_{k} = \sqrt{\frac{2}{N}}  \sum_{n=0}^{N-1} \vecz_{n}\supp{RF} e^{-j2\pi (k/N +
			f\sub{c}/f\sub{s})n}, \quad k\in \setS.
\end{equation}
Similarly, let
\begin{equation}\label{eq:noise-bb}
	\hat{\vecd}_{k} = \sqrt{\frac{2}{N}}  \sum_{n=0}^{N-1}
	\vecd_{n} e^{-j2\pi (k/N +
			f\sub{c}/f\sub{s})n}
\end{equation}
and
\begin{equation}\label{eq:quantization-error-bb}
	\hat{\vece}_{k} = \sqrt{\frac{2}{N}}  \sum_{n=0}^{N-1} \vece_{n} e^{-j2\pi (k/N +
			f\sub{c}/f\sub{s})n}.
\end{equation}
It follows from Bussgang's
decomposition~\eqref{eq:bussgang-rf}, together
with~\eqref{eq:RF-nonquantized-output}, and~\eqref{eq:received-signal-frequency-domain},
that
\begin{equation}\label{eq:io-relation-bussgang-bb}
	\hat{\vecz}_{k}\supp{BB} = \matG (\widehat{\matH}_{k}\hat{\vecs}_{k} + \hat{\vecw}_{k} + \hat{\vecd}_{k}) +
	\hat{\vece}_{k}, \quad k\in \setS.
\end{equation}
%
%

It will turn out convenient to characterize the covariance matrix
$\matC_{\hat{\vece}_{k}}=\Ex{}{\hat{\vece}_{k}\herm{(\hat{\vece}_{k})}}$.
It follows from~\eqref{eq:quantization-error-bb} that
\begin{equation}\label{eq:covariance-quantization-error-rf}
	\matC_{\hat{\vece}_{k}} = 2 \sum_{m=0}^{N-1}  \matR_{\vece}[m]  e^{-j 2\pi (k/N + f\sub{c}/f\sub{s})m},
\end{equation}
where $\matR_{\vece}[m]$ can be evaluated in closed form by
substituting~\eqref{eq:matrix-R-q-RF},~\eqref{eq:autocorrelation-transmitted-signal},
and~\eqref{eq:arcsine-law} into~\eqref{eq:covariance_quantization_error_RF}.

\section{Analysis of the \gls{evm}}
We assume that the \gls{cpu} uses its knowledge of the channel $\widehat{\matH}_{k}$,
$k\in\setS$, to recover the transmitted
signal in the frequency domain, $\hat{\vecs}_k$, via linear combining.
Specifically, consider an arbitrary frequency-dependent linear combiner, and
denote by  $\matA_k \in \complexset^{U\times B}$, the combining matrix for the
$k$th frequency sample.
We assume that the \gls{cpu} obtains an estimate $\hat{\vecs}\supp{est}_k \in \complexset^U$ of $\hat{\vecs}_k$ as
\begin{equation}\label{eq:symbol-estimate}
	\hat{\vecs}\supp{est}_k = \matA_k \hat{\vecz}_{k}\supp{BB} = \matA_k\bigl(  \matG (\widehat{\matH}_{k}\hat{\vecs}_{k} + \hat{\vecw}_{k} + \hat{\vecd}_{k}) +
	\hat{\vece}_{k}\bigr).
\end{equation}

We characterize the quality of the recovered signal in terms of the average \gls{evm} $\eta$, which we
define as\footnote{This quantity is sometimes referred to also as modulation
	error ratio.}
\begin{equation}\label{eq:evm}
	\eta = \sqrt{\frac{\sum_{k\in \setS}  \Ex{}{\vecnorm{\hat{\vecs}\supp{est}_k- \hat{\vecs}_k}^2}  }{\sum_{k\in \setS} \Ex{}{\vecnorm{\hat{\vecs}_k}^2 }}} =
	\sqrt{\frac{\sum_{k\in \setS}  \Ex{}{\vecnorm{\hat{\vecs}\supp{est}_k-
					\hat{\vecs}_k}^2}  }{E\sub{s}US}},
\end{equation}
where in the last step we used that the entries of $\hat{\vecs}_k$
have zero mean and variance $E\sub{s}=\bar{E}\sub{s}/B$.

Clearly, among all possible linear spatial combiners, {the Bussgang \gls{lmmse} combiner~\cite{bjornson19-02b,kolomvakis20-06a,nguyen21-11a}}, i.e., the \gls{lmmse} combiner based on the
Bussgang decomposition~\eqref{eq:symbol-estimate}, minimizes the \gls{evm}.
However, this combiner may not always be implementable because of complexity constraints,
or lack of knowledge of second-order statistics of the channel.
Hence, in the remainder of the section, we will also provide expressions for the \gls{evm}
obtainable using lower-complexity combiners such as
	{the Bussgang \gls{mr} and
		Bussgang \gls{zf} combiners~\cite{jacobsson18-06a,nguyen21-11a}}.
{The interested reader can find an explicit derivation of all these combiners in, e.g.,~\cite[Sec. II.C]{nguyen21-11a}.}
\subsubsection{Bussgang \gls{mr} Combiner} 
\label{sec:maximum-ratio-combiner}
We assume that the \gls{cpu} uses a Bussgang \gls{mr} combiner.
Specifically, in view of~\eqref{eq:symbol-estimate}, we set\footnote{In the
	remainder of the paper, the index $k$ is always assumed to belong to the set
	$\setS$, although this may not be explicitly mentioned.}
\begin{equation}\label{eq:MR-combiner}
	\matA_k = \matE_k \matG^{-1},
\end{equation}
where\footnote{The expression for the combiners in~\eqref{eq:MR-combiner} and~\eqref{eq:zero-forcing-combiner} include
	an array-gain normalization factor.}
\begin{equation}\label{eq:matEk}
	\matE_k = \diag\lefto(\herm{\widehat{\matH}}_k {\widehat{\matH}_k}\right)^{-1}\herm{\widehat{\matH}}_k.
\end{equation}
Let now
\begin{equation}\label{eq:matBk}
	\matB_k =\diag\lefto(\herm{\widehat{\matH}}_k
	{\widehat{\matH}_k}\right)^{-1} \left(\herm{\widehat{\matH}}_k{\widehat{\matH}_k}
	- \diag\lefto(\herm{\widehat{\matH}}_k {\widehat{\matH}_k}\right)\right).
\end{equation}
Then, the estimated sample at the \gls{cpu} after linear combining can be expressed as
\begin{equation}\label{eq:symbol-estimate-mr}
	\hat{\vecs}_k\supp{est} = \hat{\vecs}_{k}  + \matB_k \hat{\vecs}_k + \matE_k \bigl(\hat{\vecw}_k + \hat{\vecd}_k
	\bigr)
	+ \matE_k \matG^{-1} \hat{\vece}_k.
\end{equation}
Substituting~\eqref{eq:symbol-estimate-mr} into~\eqref{eq:evm}, and using that
$\hat{\vece}_k$ is uncorrelated with all other quantities, we obtain
\begin{IEEEeqnarray}{rCl}
	\eta^2 &=& \frac{1}{E\sub{s}US}  \sum_{k\in \setS} \biggl(
	E\sub{s} \Ex{}{\tr\lefto( \matB_k\herm{\matB_k} \right)}\notag\\
	&&+ (N_0+E\sub{d}) \Ex{}{\tr\lefto(\matE_k\herm{\matE_k}\right)}\\
	&&+ \Ex{}{\tr\mathopen{}\left( \matE_k\matG^{-1} \matC_{\hat{\vece}_k}\matG^{-1}\herm{\matE_k} \right)}
	\biggr).\label{eq:evm-mr}
\end{IEEEeqnarray}
\subsubsection{Bussgang \gls{zf} Combiner} 
\label{sec:Zero-forcing-combiner}
We consider now the case in which a Bussgang \gls{zf} combiner is used.
Specifically, we assume that
\begin{equation}\label{eq:zero-forcing-combiner}
	\matA_k = \matF_k \matG^{-1},
\end{equation}
where
\begin{equation}\label{eq:matFk}
	\matF_k = \left(\herm{\widehat{\matH}}_k\widehat{\matH}_k\right)^{-1}
	\herm{\widehat{\matH}_k}.
\end{equation}
Then the sample estimate $\hat{\vecs}_k\supp{est}$ at the \gls{cpu} after linear combining can be expressed as
\begin{equation}\label{eq:symbol-estimate-zf}
	\hat{\vecs}_k\supp{est} = \hat{\vecs}_{k}  + \matF_k \bigl(\hat{\vecw}_{k} +\hat{\vecd}_{k} \bigr)
	+ \matF_k \matG^{-1} \hat{\vece}_k.
\end{equation}
Substituting~\eqref{eq:symbol-estimate-zf} into~\eqref{eq:evm}, and using again that
$\hat{\vece}_k$ is uncorrelated with all other quantities, we obtain
\begin{multline}\label{eq:evm-zf}
	\eta^2 = \frac{1}{E\sub{s}US}  \sum_{k\in \setS} \biggl(
	(N_0+E\sub{d}) \Ex{}{\tr\lefto(\matF_k\herm{\matF_k}\right)} \\
	+ \Ex{}{\tr\mathopen{}\left( \matF_k\matG^{-1} \matC_{\hat{\vece}_k}\matG^{-1}\herm{\matF_k} \right)}
	\biggr).
\end{multline}
%

\subsubsection{Bussgang \gls{lmmse} Combiner} 
\label{sec:minimum-mean-square-error-combiner}
Finally, we consider the Bussgang \gls{lmmse} combiner
\begin{multline}
	\matA_k =\\  \herm{\widehat\matH_k}\left(\widehat\matH_k
	\herm{\widehat\matH_k}+ \frac{
		N_0 +E\sub{d}}{E\sub{s}}\matI_B + \frac{1}{E\sub{s}}
	\matG^{-1}\matC_{\hat{\vece}_{k}}\matG^{-1}\right)^{-1}\matG^{-1}
\end{multline}
which, as already mentioned, minimizes the \gls{evm} in~\eqref{eq:evm} and results in
\begin{IEEEeqnarray}{rCl}
	&&\Ex{}{\vecnorm{\hat{\vecs}_k\supp{est} - \hat{\vecs}_k }^2}\notag\\ &=&
	\tr\biggl(
	\Ex{}{\hat{\vecs}_k \herm{\hat{\vecs}}_k} \notag \\
	&&-
	\Ex{}{\hat{\vecs}_k \herm{(\hat{\vecz}_k\supp{BB})}}
	\Ex{}{\hat{\vecz}_k\supp{BB} \herm{(\hat{\vecz}_k\supp{BB})}}^{-1}
	\Ex{}{  \hat{\vecz}_k\supp{BB}\herm{\hat{\vecs}_k}}
	\biggr)\\
	&=& E\sub{s}U \notag\\
	&&-\tr\biggl(
	E^2\sub{s}\Exop\biggl[ \herm{\widehat{\matH}_k}
			\Bigl( E\sub{s} \widehat{\matH}_k \herm{\widehat{\matH}_k} \notag\\
			&&+ (N_0+E\sub{d}) \matI_B
			+ \matG^{-1} \matC_{\hat{\vece}_k}\matG^{-1}\Bigr)^{-1} \widehat{\matH}_k
			\biggr]
	\biggr).\IEEEeqnarraynumspace \label{eq:est-error-mmse}
\end{IEEEeqnarray}
Let now
\begin{equation}\label{eq:aux-matZ}
	\matZ_k = \herm{\widehat{\matH}_k} \left(\frac{N_0+E\sub{d}}{E\sub{s}} \matI_B +
	\frac{1}{E\sub{s}}\matG^{-1}\matC_{\hat{\vece}_k}\matG^{-1}  \right)^{-1} \widehat{\matH}_k.
\end{equation}
Substituting~\eqref{eq:est-error-mmse} into~\eqref{eq:evm} and using the matrix
inversion lemma~\cite[Sec.~0.7.4]{horn85a},
we obtain
\begin{IEEEeqnarray}{rCl}
	\eta^2 &=& 1 - \frac{1}{US}\sum_{k\in \setS}   \tr\lefto(\Ex{}{\matZ_k - \matZ_k(\matI_U + \matZ_k)^{-1}\matZ_k} \right)  \\
	&=&  \frac{1}{US} \sum_{k\in\setS} \tr\lefto( \Ex{}{(\matI_U +\matZ_k)^{-1}}
	\right).\label{eq:evm-mmse}
\end{IEEEeqnarray}

Please note that the expectations in~\eqref{eq:evm-mr},~\eqref{eq:evm-zf}, and~\eqref{eq:evm-mmse} are only with respect to the distribution of the fading channels.
The expectations with respect to the input symbols, the dither signal, and the additive noise are
evaluated in closed form.

We provide next insights on the tradeoff between \gls{osr} and number of \glspl{ap}, for a given
fronthaul constraint, and on the impact of the fronthaul constraint on
performance
by focusing on the special case of flat-fading channel and
considering the asymptotic limit $N\to\infty$.

\section{Asymptotic Characterization of the \gls{evm}}\label{sec:asympotic}
We now provide an asymptotic characterization of the \gls{evm} for the case in which the
channel between each \gls{ue} and \gls{ap} is frequency flat, i.e., $\widehat{\matH}_{k}={\matH}$ for all $k\in \setS$.
Specifically, for a fixed transmitted-signal bandwidth $W$, and a fixed number of
\glspl{ap} $B$, we consider the asymptotic regime in which the sampling rate $f\sub{s}$ goes to infinity.
This is equivalent to assuming that the number of samples $N=Tf\sub{s}$ over the observation interval $T$,
the \gls{osr} $O=N/S$ (where $S=WT)$, and the fronthaul rate $R\sub{fh}=Bf\sub{s}$ all go to infinity.
Our main finding is that, in this asymptotic limit, the Bussgang gain matrix $\matG$ and the covariance matrix $\matC_{\hat{\vece}_{k}}$ of the quantization
error both converge to a scaled identity matrix, provided that
$E\sub{d}>0$.
This allows us to replace~\eqref{eq:io-relation-bussgang-bb} with an equivalent input-output relation in which the total
additive noise is uncorrelated.
As we shall show, this equivalent input-output relation can be interpreted as the one of a conventional distributed \gls{mimo} system, in
which the \gls{cpu} has access to infinite-precision samples of the \gls{bb} signal received at each \gls{ap}.

\subsection{The Matrices $\matG$ and $\matC_{\hat{\vece}_{k}}$ in the Frequency-Flat Fading Case, and their Asymptotic Limit}
Assume that $\widehat{\matH}_{k}={\matH}$ for all $k\in \setS$ (frequency-flat fading assumption) and let
$\tp{\vech_{b}}$ denote the $b$th row of $\matH$.
Let us also introduce the following quantities:
\begin{equation}\label{eq:pu}
	p_{b} = \sqrt{\frac{S}{N}\big(E\sub{s}\vecnorm{\vech_{b}}^{2}+N_0\big)+
		\frac{E\sub{d}}{2}},\quad b=1,\dots,B,
\end{equation}
\begin{equation}\label{eq:cm}
	c[m] = \sum_{k\in \setS}e^{-j2\pi \frac{k}{N}m}  =  1+ 2\sum_{k=1}^{(S-1)/2}
	\cos(2\pi (mk)/N) \, ,
\end{equation}
where the last equality in~\eqref{eq:cm} follows from~\eqref{eq:setS} and because $S$ is odd.
Furthermore, let
\begin{equation}
	v_{bb'}[m]=  \Re\lefto\{ \bigl(E\sub{s}\tp{\vech_{b}}\conj{\vech_{b'}}+ N_{0}\bigr)
	e^{j2\pi \frac{f_{c}}{f_{s}}m}\right\}
\end{equation}

\begin{equation}\label{eq:buv}
	s_{bb'}[m] = \frac{\frac{c[m]}{N} v_{bb'}[m] +\frac{E\sub{d}}{2}\delta[m]\delta[b-b']
	}{p_{b}p_{b'}  }
\end{equation}
and
\begin{equation}\label{eq:ruv-def}
	r_{bb'}[m] = \arcsin(s_{bb'}[m]) - s_{bb'}[m],
\end{equation}
where the last three quantities are defined for all $b \in \{1,\dots,B\}$, $b' \in \{1,\dots,B\}$.
As a consequence of~\eqref{eq:bussgang-matrix},~\eqref{eq:matrix-R-q-RF},
and~\eqref{eq:autocorrelation-transmitted-signal}, we can express the entry in position $(b,b)$ of the diagonal matrix
$\matG$ as
\begin{equation}\label{eq:matG-alternative}
	[\matG]_{b,b} = \sqrt{\frac{2}{\pi}} \frac{1}{p_{b}}.
\end{equation}
Furthermore, it follows
from~\eqref{eq:covariance-quantization-error-rf},~\eqref{eq:covariance_quantization_error_RF},~\eqref{eq:autocorrelation-transmitted-signal},
and~\eqref{eq:arcsine-law} that the entry in position $(b,b')$ of the matrix $\matC_{\hat{\vece}_{k}}$ can be expressed
as
\begin{equation}\label{eq:matCe-alternative}
	\bigl[\matC_{\hat{\vece}_{k}}\bigr]_{b,b'} = \frac{4}{\pi} \sum_{m=0}^{N-1}  r_{bb'}[m]  e^{-j 2\pi (k/N +
			f\sub{c}/f\sub{s})m}.
\end{equation}

We are interested in determining the asymptotic behavior of the matrices $\matG$ and $\matC_{\hat{\vece}_{k}}$ in the
asymptotic limit $N\to\infty$.
It follows immediately from~\eqref{eq:matG-alternative} that, whenever
$E\sub{d}>0$,
\begin{equation}\label{eq:limit-matG}
	[\matG]_{b,b} \overset{\text{a.s.}}{\longrightarrow} \sqrt{\frac{4}{\pi E\sub{d}}}, \quad N\to\infty.
\end{equation}
To establish a similar result for $\matC_{\hat{\vece}_{k}}$, we provide in the following lemma a characterization of the
asymptotic behavior of $r_{bb'}[m]$ in~\eqref{eq:ruv-def}.
\begin{lem}\label{lem:ruv}
	The random variable $r_{bb'}[m]$ in~\eqref{eq:ruv-def} satisfies the
	following properties:
	\begin{equation}\label{eq:ruu0}
		r_{bb'}[m]\leq r_{bb}[0] = \frac{\pi}{2}-1
	\end{equation}
	and, for $b\neq b'$ or $m\neq 0$
	\begin{equation}\label{eq:ruvm}
		\abs{r_{bb'}[m]} \overset{\text{a.s.}}{\longrightarrow} 0, \quad N\to\infty.
	\end{equation}
	Furthermore, for all $b,b'$,
	\begin{equation}\label{eq:sum-ruvm}
		\sum_{m=1}^{N-1} \abs{r_{bb'}[m]}\overset{\text{a.s.}}{\longrightarrow} 0, \quad N\to\infty.
	\end{equation}

\end{lem}
\begin{IEEEproof}
	See Appendix~\ref{app:lem}.
\end{IEEEproof}

It now follows from Lemma~\ref{lem:ruv} and~\eqref{eq:matCe-alternative} that
\begin{equation}\label{eq:limit-entry-Ce}
	\bigl[\matC_{\hat{\vece}_{k}}\bigr]_{b,b'} \overset{\text{a.s.}}{\longrightarrow} 2\left( 1 -\frac{2}{\pi}\right)
	\delta[b-b'],\quad N\to\infty.
\end{equation}
As a consequence,
\begin{equation}\label{eq:limit-Ce}
	\matC_{\hat{\vece}_{k}} \overset{\text{a.s.}}{\longrightarrow} 2\left( 1 -\frac{2}{\pi}\right) \matI_{B}, \quad
	N\to\infty.
\end{equation}

Intuitively speaking,~\eqref{eq:limit-matG} and~\eqref{eq:limit-Ce}, imply that, in the limit $N\to\infty$, we can
replace the \gls{bb} input-output relation~\eqref{eq:io-relation-bussgang-bb} with the following input-output relation
\begin{equation}\label{eq:io-alternative-asymptotic}
	\hat{\vecz}_{k}\supp{BB} = \widehat{\matH}_{k}\hat{\vecs}_{k} +
	\tilde{\vecw}_{k}, \quad k\in \setS,
\end{equation}
where the noise $\tilde{\vecw}_{k}$, which is uncorrelated with $\hat{\vecs}_{k}$, has uncorrelated entries.
Specifically, $\Ex{}{\tilde{\vecw}_{k}\herm{\tilde{\vecw}_{k}}}= (N_{0} + \frac{\pi}{2}E\sub{d}) \matI_{B}$.
Note that the asymptotically equivalent input-output relation~\eqref{eq:io-alternative-asymptotic} coincides with that
of a conventional distributed \gls{mimo} architecture, where down-conversion and
sampling are performed at the \gls{ap}, and the (infinite-precision) samples of
the \gls{bb} signal are conveyed to the \gls{cpu} for digital processing.

The asymptotic limits~\eqref{eq:limit-matG} and~\eqref{eq:limit-Ce} cannot be directly used
in~\eqref{eq:evm-mr},~\eqref{eq:evm-zf}, and~\eqref{eq:evm-mmse} to obtain asymptotic expressions for the
\gls{evm} $\eta$ in the limit $N\to\infty$.
Indeed, one needs to first verify that the expectations in~\eqref{eq:evm-mr},~\eqref{eq:evm-zf}, and~\eqref{eq:evm-mmse}
and the limit $N\to\infty$ can be interchanged.
This is typically possible under mild assumptions on the fading distribution.
Since the exact conditions under which the interchange is possible are typically cumbersome to derive,
we provide them in the next theorem only for the case $U=1$, for which the matrix $\matH$ reduces to the $B$-dimensional
vector $\vech$.
Similar conditions can be established for the case $U>1$.
\begin{thm}\label{thm:as-evm}
	Let $f(x)=\arcsin(x)-x$ and $\gamma_{b}= \sqrt{S(E\sub{s}\abs{h_{b} }^{2}+ N_{0})
			+
			\frac{E\sub{d} }{2}}$. 
	For every fading channel distribution satisfying the technical
	conditions
	\begin{equation}\label{eq:technical-cond-mr}
		\sum_{b=1}^{B}\sum_{b'=1}^{B}
		\Ex{}{\frac{h^{*}_{b}h_{b'}}{\vecnorm{\vech}^{4}} \gamma_{b}\gamma_{b'}} <\infty
	\end{equation}
	and
	\begin{equation}\label{eq:technical-cond-mr-2}
		\sum_{b=1}^{B}\sum_{b'=1}^{B}
		\Ex{}{\frac{h^{*}_{b}h_{b'}}{\vecnorm{\vech}^{4}} \gamma_{b}\gamma_{b'}
		\left(\frac{2SE_{s}\abs{h_{b}} \abs{h_{b'}} +N_{0}}{E\sub{d}}\right)}
		<\infty,
	\end{equation}
	the \gls{evm} $\eta$ with Bussgang \gls{mr}/\gls{zf} combining for the case $U=1$
	admits the following asymptotic characterization:
	\begin{equation}\label{eq:evm-mr-as}
		\lim_{N \to \infty} \eta^{2} =  \frac{N_{0} + (\pi/2)E\sub{d}}{E\sub{s} }
		\Ex{}{\frac{1}{\vecnorm{\vech}^{2}}}.
	\end{equation}
	Furthermore, let
	\begin{equation}\label{eq:def-tbb}
		t_{bb'} = \left( \frac{\pi}{2}-1 \right) \gamma_{b}\gamma_{b'}
	\end{equation}
	and
	\begin{equation}\label{eq:def-tbb'}
		\bar{t}_{bb'} = \left( \frac{\pi}{2}-1 \right)S \gamma_{b}\gamma_{b'}
		\left(\frac{2SE_{s}\abs{h_{b}} \abs{h_{b'}} +N_{0}}{E\sub{d}}\right).
	\end{equation}
	For every channel law satisfying the technical condition
	\begin{equation}\label{eq:technical-cond-mmse}
		\Ex{}{\frac{1}{1+\bar{\gamma}}}<\infty,
	\end{equation}
	where
	\begin{multline}\label{eq:bar-gamma}
		\bar{\gamma}=  \vecnorm{\vech}^{2} \biggl(  \sum_{b=1}^{B}\gamma_{b}^{2}
		\left( \frac{\pi}{2}-1 \right)+ \frac{2}{E\sub{s} }
		\sum_{b=1}^{B}\sum_{b'=1}^{B} t_{bb'}\\ +
		\frac{2}{E\sub{s}S}\sum_{b=1}^{b}\sum_{b'=1}^{B} \bar{t}_{bb'} \biggr)^{-1},
	\end{multline}
	the \gls{evm} with Bussgang \gls{lmmse} combining
	admits the following asymptotic characterization:
	\begin{equation}\label{eq:evm-mmse-as}
		\lim_{N\to\infty} \eta^{2} = \Ex{}{\left(1+ \frac{E\sub{s} }{N_{0} +
				(\pi/2) E\sub{d}
			}\vecnorm{\vech}^{2}\right)^{-1} }.
	\end{equation}

\end{thm}
\begin{IEEEproof}
	See Appendix~\ref{app:thm}.
\end{IEEEproof}
\subsection{Insights from the Asymptotic \gls{evm}
	Characterization}\label{sec:insights}

Some remarks on  Theorem~\ref{thm:as-evm} are in order.
The \gls{evm} expressions~\eqref{eq:evm-mr-as} and~\eqref{eq:evm-mmse-as}
illustrate that, for a given $E\sub{d}>0$,
the in-band interference caused by the nonsubtractive, wideband dither signal causes a
deterioration of the SNR of the equivalent conventional distributed \gls{mimo}
architecture from $E\sub{s}/N_{0}$ to $E\sub{s}/(N_{0} + (\pi/2)E\sub{d})$.
The term $\pi/2$ stems from our assumption that, as $N$ grows, the bandwidth of
the dither signal increases, and is fundamental in wideband-regime analyses
of $1$-bit quantization systems (see, e.g.,~\cite{koch13-09a}).

It is worth noting, though, that~\eqref{eq:evm-mr-as} and~\eqref{eq:evm-mmse-as}
hold \emph{for every} $E\sub{d}>0$.
This implies that the asymptotic \gls{evm} with Bussgang \gls{mr}/\gls{zf}
can be made arbitrarily close to
\begin{equation}\label{eq:evm-mr-as-no-dither}
	\frac{N_{0}}{E\sub{s}} \Ex{}{\frac{1}{\vecnorm{\vech}^{2}}}.
\end{equation}
Similarly, the asymptotic \gls{evm} with Bussgang \gls{lmmse} can be made
arbitrarily close to
\begin{equation}\label{eq:evm-mmse-as-no-dither}
	\Ex{}{\left(1+\frac{E\sub{s}}{N_{0}}\vecnorm{\vech}^{2}\right)^{-1}}.
\end{equation}
Note that~\eqref{eq:evm-mr-as-no-dither} and~\eqref{eq:evm-mmse-as-no-dither}
are the squared \gls{evm} values achievable with \gls{mr}/\gls{zf} and \gls{lmmse}
combiners, respectively, in a conventional homodyne distributed \gls{mimo} architecture, with
infinite precision converters.
%
In other words, if we allow for~$N$ to be sufficiently large, and we set
$E\sub{d}$ sufficiently small, the nonlinear distortion caused by the $1$-bit quantization
in the architecture depicted in Fig.~\ref{fig:system}, has no impact on
performance.

The proofs of Lemma~\ref{lem:ruv} and Theorem~\ref{thm:as-evm} provide also
relevant insights into the optimal choice of $E\sub{d}$ for the practically
relevant case of finite~$N$,
and shed light on the dependence of $E\sub{d}$ on
system parameters such as the fronthaul rate, the number of \glspl{ap}, and the type of combiner used at the \gls{cpu}.

Indeed, note that the results reported in Lemma~\ref{lem:ruv}, and, hence,
also~\eqref{eq:limit-Ce}, hold also if we fix~$N$ and let $E\sub{d}\to\infty$.
This implies that, for every fixed $N$, we can whiten the correlation matrix $\matC_{\hat{\vece}_{k}}$ of
the quantization error, simply by increasing $E\sub{d}$.
This, however, causes an increase of the in-band interference caused by the
dither signal, which, for sufficiently large $N$, yields the
$(\pi/2)E\sub{d}$ penalty term in~\eqref{eq:evm-mr-as}
and~\eqref{eq:evm-mmse-as}.
So for a finite $N$, we expect that there will be a tension between increasing
$E\sub{d}$ to whiten the quantization noise, and reducing $E\sub{d}$ to limit
in-band interference.

Some insights into this tension can be obtained by analyzing the term $p_b$ in~\eqref{eq:pu}.
For the quantization noise to whiten, we need that $p^{2}_{b}\approx
	E\sub{d}/2$ for all $b=1,\dots,B$, which occurs whenever
\begin{equation}\label{eq:whitening-condition}
	\frac{E\sub{d}}{2N_{0}}\gg \frac{S}{N}
	\left(\frac{E\sub{s}}{N_{0}}\abs{h_{b}}^{2} +1\right).
\end{equation}
At the same time, we would like to minimize $E\sub{d}$, to reduce the impact of
in-band interference caused by the dither signal.
We see from~\eqref{eq:whitening-condition} that the higher the \gls{osr}
$N/S$ or, equivalently, the higher the fronthaul rate $R
	\sub{fh}$, for a fixed number of \glspl{ap} $B$, the smaller the value of $E\sub{d}$ necessary
for~\eqref{eq:whitening-condition} to hold.
The dependence of $E\sub{d}$ on $B$ for a fixed \gls{osr} $N/S$ is more intricate.
On the one hand, the larger $B$, the smaller $E\sub{s}$ for a fixed normalized
energy $\bar{E}\sub{s}$, because of the energy-efficient
assumption~\eqref{eq:energy-efficient-setup}.
This implies that a smaller $E\sub{d}$ is needed to
satisfy~\eqref{eq:whitening-condition} for a fixed $\abs{h_{b}}^{2}$.
On the other hand, the larger $B$, the smaller the distance between each
\gls{ue} and the closest \gls{ap},
which results in a higher average $\abs{h_{b}}^{2}$ for that \gls{ap}.
This, in turn, results in a larger $E\sub{d}$ to
satisfy~\eqref{eq:whitening-condition}.
Also, how much the left-hand-side of~\eqref{eq:whitening-condition} should
exceed the right-hand side of~\eqref{eq:whitening-condition} depends on the combiner used at the \gls{cpu}.
We expect that Bussgang \gls{lmmse} will
be, in general, less sensitive to $E\sub{d}$,
because it exploits the knowledge of the covariance matrix of the quantization
noise.
Note that, although we focused this discussion on the case $U=1$, the insights we
have just reported hold for arbitrary $U$, since they rely on the convergence
result~\eqref{eq:limit-Ce}, which we have established for arbitrary~$U$.

\section{Numerical Investigations}\label{sec:numerical-investigations}
\subsection{Single UE}\label{sec:single-ue}
\subsubsection{Scenario}\label{sec:scenario}
Throughout this subsection, we focus on the case $U=1$ and investigate
the impact on the \gls{evm} of the fronthaul
constraint~\eqref{eq:fronthaul-constraint}, the number of \glspl{ap} deployed within a given coverage
area, and the energy of the dither signal.

\begin{figure}
	\subcaptionbox{$U=1$\label{fig:topology-u1}}[0.45\textwidth]{\includegraphics[width=0.45\textwidth]{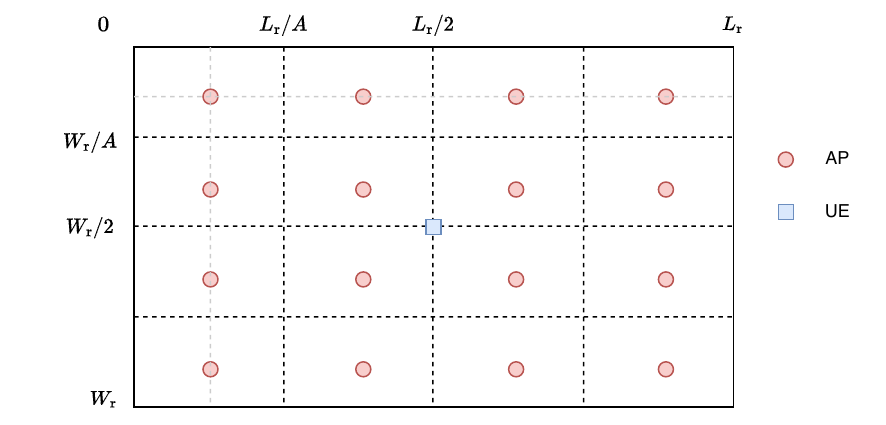}}
	\subcaptionbox{$U=4$\label{fig:fix4uetopology}}[0.45\textwidth]{\includegraphics[width=0.45\textwidth]{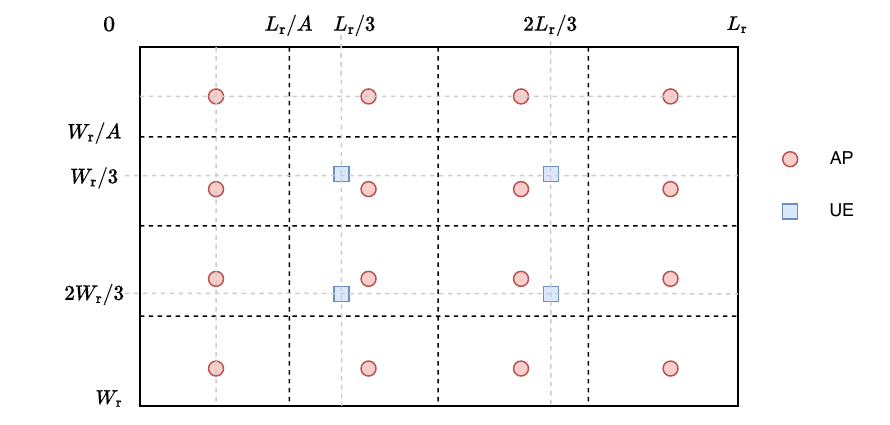}}
	\caption{The topology of the distributed \gls{mimo} system for the case $U=1$ and $U=4$
		and fixed \gls{ue} positions considered
		in this section. In the two figures,
		$B=16$.}\label{fig:topology}
\end{figure}

We consider the scenario depicted in Fig.~\ref{fig:topology-u1}: a rectangular
area of size $L\sub{r}\times W\sub{r}$ is covered uniformly by $B=A^{2}$
\glspl{ap}, with  $A$ chosen to be an even number.
The \gls{ue} is placed at the center of the coverage area.
In the remainder of the paper, we will assume that
$L\sub{r}=W\sub{r}=\qty{100}{\metre}$.
We also assume that the \glspl{ap} are at height of $\qty{10}{\metre}$ and that the \gls{ue} is
at height of $\qty{0}{\metre}$.
Furthermore, the $B$-dimensional (frequency-flat) channel vector $\vech$, whose entries describe
the random propagation channel between the \gls{ue} and each \gls{ap}, is assumed
to follow a
$\jpg(\veczero_{B},\matL)$ distribution, where the diagonal matrix~$\matL$ contains
on its diagonal the path-loss coefficients between the \gls{ue} and each \gls{ap}.
The path-loss coefficient, measured in \unit{\decibel}, between the \gls{ue} and the $b$th \gls{ap}
is modeled as
\begin{equation}\label{eq:path-loss}
	10\log_{10} [\matL]_{b,b} = -37.6 \log_{10} (d_{b}/d_{0}) - 35.3\, ,
\end{equation}
where $d_{b}$ is the distance in meters between the \gls{ue} and the $b$th
\gls{ap}, and $d_{0}=\qty{1}{m}$.
In this section, we shall also consider a noise spectral density $N_{0}$ that
results in a noise power of
$\qty{-94}{dBm}$ and a normalized energy per sample  $\bar{E}\sub{s}$, defined
in~\eqref{eq:energy-efficient-setup}, of $\qty{20}{dBm}$.
We also set $f\sub{c}=\qty{2.4}{GHz}$, $W=\qty{24}{MHz}$, and $S=9$, and assume
a frequency-flat fading.
\subsubsection{Impact of Dithering}
\begin{figure}
	\subcaptionbox{Bussgang \gls{mr}/\gls{zf}
		combiner\label{fig:dither_zf_u1}}[0.5\textwidth]{\includegraphics[width=0.4\textwidth]{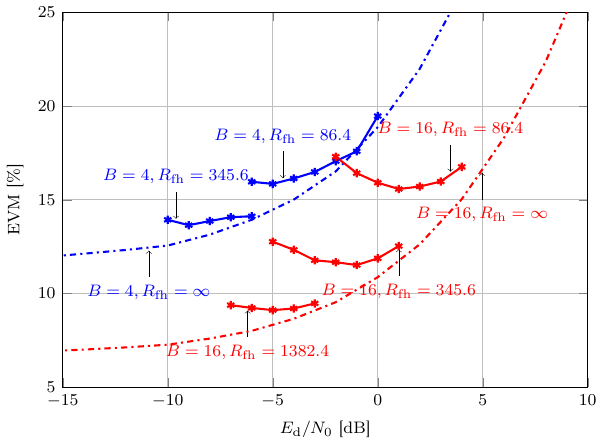}}
	\subcaptionbox{Bussgang \gls{lmmse}
		combiner\label{fig:dither_mmse_u1}}[0.5\textwidth]{\includegraphics[width=0.4\textwidth]{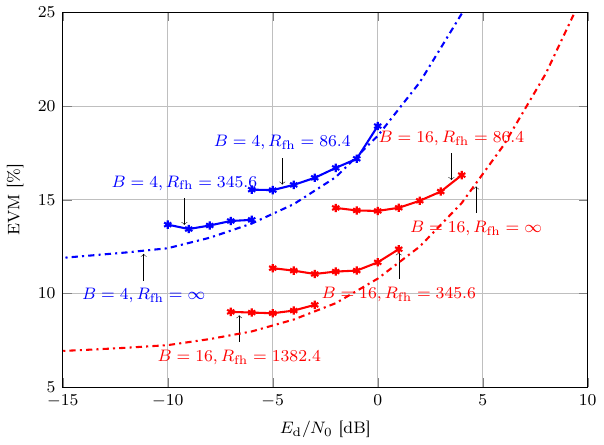}}
	\caption{\gls{evm} as a function of the dither-to-noise ratio $E\sub{d}/N_{0}$, for the case
	of Bussgang \gls{mr}/\gls{zf} and \gls{lmmse} combiners and $U=1$.
	The values of $R\sub{fh}$ are given in \unit{Gbit/s}.
	The dot-dashed lines denote the
	\gls{evm} for the case of infinite fronthaul rate, computed
	using~\eqref{eq:evm-mr-as} and~\eqref{eq:evm-mmse-as}, respectively.
	}\label{fig:u1-dither}
\end{figure}
To validate the insights reported in Section~\ref{sec:insights}, based on the
asymptotic limit $N\to\infty$, we evaluate in this section the impact of the
dither-to-noise ratio $E\sub{d}/N_{0}$ on the \gls{evm}.
In Fig.~\ref{fig:u1-dither}, we report the \gls{evm} as a function of
$E\sub{d}/N_{0}$ for $B\in \{4,16\}$ and $R\sub{fh}\in \{\qty{86.4}{Gbit/s},
	\qty{345.6}{Gbit/s}, \qty{1382.4}{Gbit/s}\}$
for both Bussgang \gls{mr}/\gls{zf} combiner and \gls{lmmse} combiner, computed using the
exact nonasymptotic \gls{evm} expressions in~\eqref{eq:evm-mr} and~\eqref{eq:evm-mmse},
as well as the asymptotic ($R\sub{fh}\to\infty$) \gls{evm} value, computed
using~\eqref{eq:evm-mr-as} and~\eqref{eq:evm-mmse-as}.\footnote{In the remainder of the paper, the
	\gls{osr} corresponding to a given pair $(B,R\sub{fh})$ is obtained by assuming
	that~\eqref{eq:fronthaul-constraint} holds with equality.}
It is worth mentioning that the testbed designed in~\cite{aabel20-11p,sezgin21-02a} involves a
\gls{cpu} that can be connected to up to $12$ \glspl{ap} and is equipped with $1$-bit \glspl{adc}
operating at $\qty{10}{GS/s}$. This yields a total fronthaul rate of $\qty{120}{Gbit/s}$, which is
within the range of fronthaul-rate values considered in this section.

In agreement with our predictions in Section~\ref{sec:insights}, for each fixed
$B$ and $R\sub{fh}$, there exists an optimal non-zero $E\sub{d}/N_{0}$ value
that minimizes the \gls{evm}.
Operating at a $E\sub{d}/N_{0}$ larger than the optimal one, allows us to
approach the asymptotic \gls{evm} predicted by~\eqref{eq:evm-mr-as} and~\eqref{eq:evm-mmse-as}
for that $E\sub{d}/N_{0}$ value, at the cost of an \gls{evm} penalty.
We also observe that, as predicted in Section~\ref{sec:insights}, for a
fixed~$B$, a larger
$R\sub{fh}$ results in a smaller $E\sub{d}/N_{0}$.
To determine the dependence of $E\sub{d}/N_{0}$ on $B$ for a fixed
\gls{osr} $O=R\sub{fh}/(BW)$, we compare the curves corresponding to $B=4,
	R\sub{fh}=\qty{86.4}{Gbit/s}$ with the curve corresponding to $B=16, R\sub{fh}=\qty{354.6}{Gbit/s}$,
since, in both cases, $O=900$.
We can also compare the curve corresponding to $B=4,
	R\sub{fh}=\qty{345.6}{Gbit/s}$ with the curve corresponding to $B=16,
	R\sub{fh}=\qty{1382.4}{Gbit/s}$, since, in both cases $O=3600$.
It turns out that in both scenarios, the optimal $E\sub{d}/N_{0}$ increases
when we increase $B$ for a fixed $O$.
This suggests that the increase in $E\sub{d}/N_{0}$ caused by the reduction of
the distance between \gls{ue} and closest \gls{ap} dominates over the
reduction in $E\sub{d}/N_{0}$ enabled by the lower transmitted power
(see~\eqref{eq:energy-efficient-setup}).
Finally, by comparing Fig.~\ref{fig:dither_zf_u1} with Fig.~\ref{fig:dither_mmse_u1},
we note, again in agreement with the observations in
Section~\ref{sec:insights}, that the value of $E\sub{d}/N_{0}$ that minimizes the
\gls{evm} is lower when the Bussgang \gls{lmmse} combiner is used.

\subsubsection{Impact of Fronthaul Rate}
\begin{figure}
	\subcaptionbox{Bussgang \gls{mr}/\gls{zf}
		combiner\label{fig:zf_u1}}[0.5\textwidth]{\includegraphics[width=0.4\textwidth]{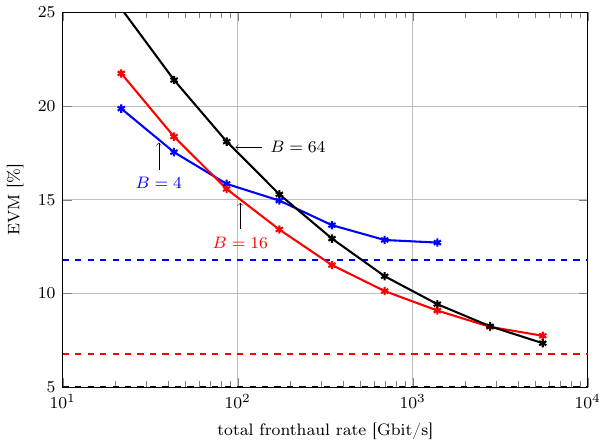}}
	\subcaptionbox{Bussgang \gls{lmmse}
		combiner\label{fig:mmse_u1}}[0.5\textwidth]{\includegraphics[width=0.4\textwidth]{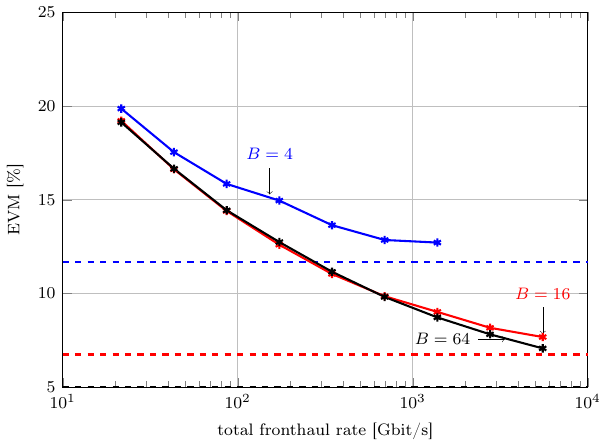}}
	\caption{\gls{evm} as a function of the fronthaul rate $R\sub{fh}$, for the case
		of Bussgang \gls{mr}/\gls{zf} and \gls{lmmse} combiners and $U=1$.
		The value of $E\sub{d}$ is optimized for each value of fronthaul rate
		considered in the figure.
		The dashed lines denote the
		\gls{evm} for the case of infinite fronthaul rate, computed
		using~\eqref{eq:evm-mr-as} and~\eqref{eq:evm-mmse-as}, and setting
		$E\sub{d}=0$, in accordance to the discussion in Section~\ref{sec:insights}.
	}\label{fig:u1-evm}
\end{figure}
Still focusing on the case $U=1$, in Fig.~\ref{fig:u1-evm}, we illustrate the dependence of the \gls{evm} on the
fronthaul rate $R\sub{fh}$ for the case of Bussgang \gls{mr}/\gls{zf} combiner
and Bussgang \gls{lmmse} combiner, computed using~\eqref{eq:evm-mr}
and~\eqref{eq:evm-mmse} (the value of $E\sub{d}$ is optimized for each value of fronthaul rate
considered in the figure), and compare the resulting nonasymptotic
\gls{evm} with the asymptotic values
in~\eqref{eq:evm-mr-as} and~\eqref{eq:evm-mmse-as}, computed by setting
$E\sub{d}=0$, in accordance to the discussion in
Section~\ref{sec:insights}.\footnote{Using~\cite[Eq.~(1)]{cressie81-08a}, the expectation in~\eqref{eq:evm-mr-as} can
	be computed in closed form for
	the channel model considered in this section.}
For the case of \gls{mr}/\gls{zf} combiner, we see that, for small values of
$R\sub{fh}$, increasing the number of \glspl{ap} is deleterious in
terms of \gls{evm}, and the smallest \gls{evm} is achieved when $B=4$.
This suggests that, when $R\sub{fh}$ is small, it is better to use the available
total fronthaul rate to achieve a large \gls{osr} (for the lowest
value of $R\sub{fh}$ considered in the figure, i.e., $\qty{21.6}{Gbit/s}$, we
have that $O=225$), rather than
to densify the \gls{ap} deployment.

As $R\sub{fh}$ increases, the \gls{evm} decreases for each fixed $B$, but at
different speeds: the larger $B$, the faster the rate of decay, in agreement
with the asymptotic results in~\eqref{eq:evm-mr-as} and~\eqref{eq:evm-mmse-as},
which, for the channel model used in this section, predict (as illustrated in
Fig.~\ref{fig:u1-evm}) that the asymptotic
\gls{evm} decreases monotonically with~$B$.

With \gls{zf}, maintaining optimality in terms of \gls{evm} requires one to
progressively switch from a deployment with less \glspl{ap} to a
deployment with more \glspl{ap} as the total fronthaul rate is increased.
Indeed, we see that we need to switch from $4$ to $16$ \glspl{ap}
when the total fronthaul rate is larger or equal to $\qty{86.4}{Gbit/s}$ and then switch again
from $16$ to $64$ when the total fronthaul rate is larger or equal to
$\qty{2764.8}{Gbit/s}$.

The trend with \gls{lmmse} is different.
Although the \gls{evm} still decays monotonically with $R\sub{fh}$, there is no
penalty in setting $B=64$ when $R\sub{fh}$ is small.
Indeed, when $R\sub{fh}$ is small, the \gls{evm} achieved for the three values
of $B$ considered in the figure are similar.
This suggests that, in this regime, one can trade spectral oversampling with
temporal oversampling.

\subsection{Multiple \glspl{ue}}
\subsubsection{Fixed \glspl{ue} Positions}
\begin{figure}
	\subcaptionbox{Bussgang \gls{zf}
		combiner\label{fig:dither_zf_u4}}[0.5\textwidth]{\includegraphics[width=0.4\textwidth]{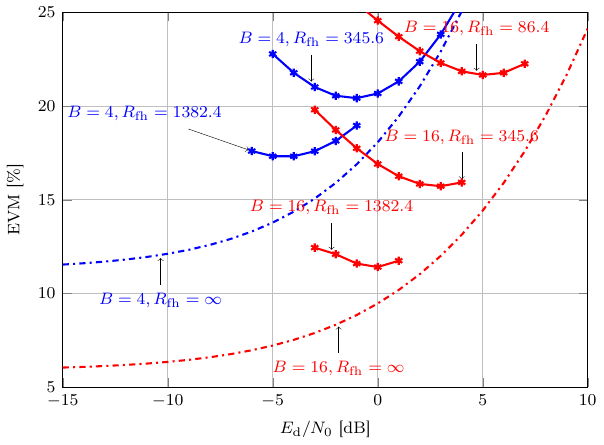}}
	\subcaptionbox{Bussgang \gls{lmmse}
		combiner\label{fig:dither_mmse_u4}}[0.5\textwidth]{\includegraphics[width=0.4\textwidth]{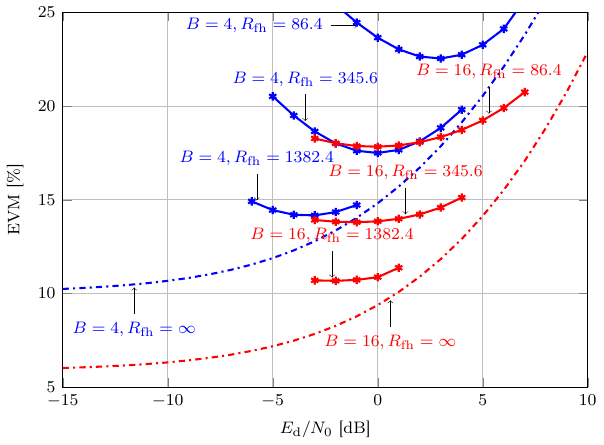}}
	\caption{\gls{evm} as a function of the dither-to-noise ratio $E\sub{d}/N_{0}$, for the case
	of Bussgang \gls{zf} and \gls{lmmse} combiners and $U=4$. The dot-dashed lines denote the
	\gls{evm} for the case of infinite fronthaul rate, computed
	by substituting~\eqref{eq:limit-matG}
	and~\eqref{eq:limit-Ce} into~\eqref{eq:evm-zf} and~\eqref{eq:evm-mmse}.
	}\label{fig:u4-dither}
\end{figure}
We next generalize the analysis performed in Section~\ref{sec:single-ue} to the
case of multiple \glspl{ue} ($U>1)$.
	We start by considering the scenario depicted in Fig.~\ref{fig:fix4uetopology},
	where $U=4$ \glspl{ue} are deterministically placed in a square grid.

	First, we verify whether the insights we reported in Section~\ref{sec:insights} hold
	also for the case $U>1$.
	In Fig.~\ref{fig:u4-dither}, we plot the \gls{evm} achievable with \gls{zf}
	and \gls{lmmse} combiners, computed using~\eqref{eq:evm-zf}
	and~\eqref{eq:evm-mmse}, for the same $(B,R\sub{fh})$ pairs we analyzed in Fig.~\ref{fig:u1-dither}.
	We also plot their asymptotic limit for $R\sub{fh}\to\infty$ obtained this
	time by substituting~\eqref{eq:limit-matG}
	and~\eqref{eq:limit-Ce} into~\eqref{eq:evm-zf}
	and~\eqref{eq:evm-mmse}.
	We do not consider the \gls{mr} combiner, since
	for the scenario considered in this section, it yields an \gls{evm} exceeding
$40\%$ for both $B=16$ and $B=4$ even in the asymptotic limit
$R\sub{fh}\to\infty$ and when $E\sub{d}=0$.
	This high \gls{evm} value is caused by residual multiuser interference.

	As shown in Fig.~\ref{fig:u4-dither}, the findings reported in
	Section~\ref{sec:insights} still hold, namely,
	\begin{inparaenum}[i)]
		\item for each $(B,R\sub{fh})$ pair, there exists a nonzero $E\sub{d}/N_0$ value that minimizes the \gls{evm};
		\item for fixed $R\sub{fh}$, the asymptotic limit is approached by increasing $E\sub{d}/N_0$ beyond the optimal value;
		\item larger fronthaul values result in smaller $E\sub{d}/N_0$;
		\item for a given \gls{osr}, increasing $B$ increases $E\sub{d}/N_0$;
		\item the nonasymptotic \gls{evm} curve is flatter around its minimum with
		\gls{lmmse} than with \gls{zf}.
	\end{inparaenum}

	\begin{figure}
		\subcaptionbox{Bussgang \gls{zf}
			combiner\label{fig:zf_u4}}[0.5\textwidth]{\includegraphics[width=0.4\textwidth]{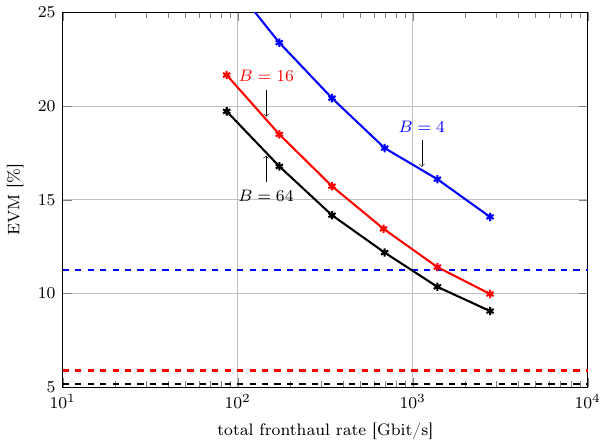}}
		\subcaptionbox{Bussgang \gls{lmmse}
			combiner\label{fig:mmse_u4}}[0.5\textwidth]{\includegraphics[width=0.4\textwidth]{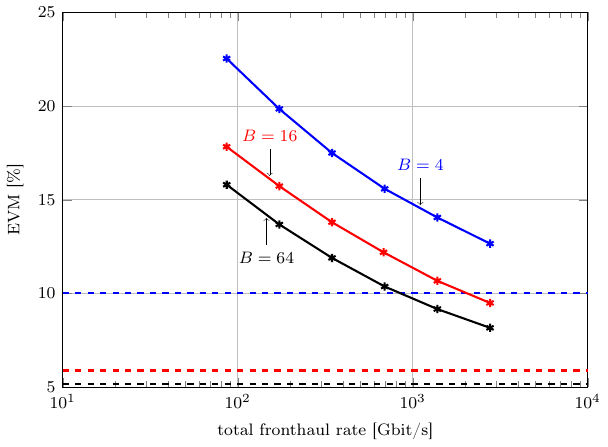}}
		\caption{\gls{evm} as a function of the fronthaul rate $R\sub{fh}$, for the case
			of Bussgang \gls{zf} and \gls{lmmse} combiners and $U=4$.
			The value of $E\sub{d}$ is optimized for each value of fronthaul rate
			considered in the figure.
			The dashed lines denote the
			\gls{evm} for the case of infinite fronthaul rate, computed substituting~\eqref{eq:limit-matG}
			and~\eqref{eq:limit-Ce} into~\eqref{eq:evm-zf} and~\eqref{eq:evm-mmse}, and setting
			$E\sub{d}=0$, in accordance to the discussion in Section~\ref{sec:insights}.
		}\label{fig:u4-evm}
	\end{figure}
	We next analyze in Fig.~\ref{fig:u4-evm} the impact of the fronthaul rate on the \gls{evm}.
	Differently for the case $U=1$ analyzed in Fig.~\ref{fig:u1-evm}, we see that,
	for both Bussgang \gls{zf} and \gls{lmmse} combiners, the architecture with
$B=64$ \glspl{ap} outperforms the architecture with $B=4$ and $B=16$ \glspl{ap}
	for all values of $R\sub{fh}$ considered in the figure.
	For the case of \gls{zf}, this means that, when $U=4$, the reduction in the distance between each \gls{ue} and the
	closest \gls{ap} overcomes the deleterious effect on the \gls{evm} of a
	reduction in the \gls{osr} when $R\sub{fh}$ is small, noticed in
	Fig.~\ref{fig:u1-evm}.
	Note also that the gap between the \gls{evm} curves and their
	asymptotic counterparts, computed substituting~\eqref{eq:limit-matG}
	and~\eqref{eq:limit-Ce} into~\eqref{eq:evm-zf} and~\eqref{eq:evm-mmse}, and setting
$E\sub{d}=0$,
	is larger than for the $U=1$ case illustrated in Fig.~\ref{fig:u1-evm},
	which suggests that larger fronthaul values are required to operate close to
	the asymptotic limits.
	\subsubsection{Random \glspl{ue} Positions}
	\begin{figure}[h]
		\centering
		\includegraphics[width=0.4\textwidth]{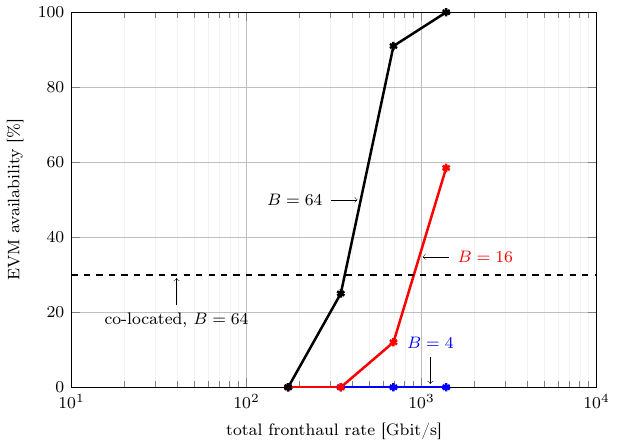}
		\caption{\gls{evm} availability for $U=4$ randomly positioned \glspl{ue}.
			The dashed line denotes the \gls{evm} availability for a co-located system in
			which a massive \gls{mimo} base station with $64$ antennas serves all four
			\glspl{ue}.}
		\label{fig:u4random}
	\end{figure}
	We consider the situation
	in which four \glspl{ue} are dropped independently and uniformly at random
	over the coverage area, while the \glspl{ap} are deployed over a square grid as
	in Fig.~\ref{fig:fix4uetopology}.
	Focusing only on the \gls{lmmse} combiner, we report in Fig.~\ref{fig:u4random} the
	\gls{evm} \emph{availability}, which we define as the probability,
	computed with respect to the random \glspl{ue} positions,
	that the \gls{evm} is smaller than $12.5\%$, as a function of the total fronthaul
	rate.
	We chose this \gls{evm} value because it is the minimum \gls{evm} required in
	LTE and 5G NR to support $16$-QAM~\cite{3gpp,5g}.
	For reference, we also illustrate the \gls{evm} achieved by a co-located massive
	\gls{mimo} architecture in which a \gls{bs}, equipped with $64$ antennas and
	positioned at the center of the coverage area, serves all $4$ \glspl{ue}. For
	this co-located architecture, we consider the favorable scenario of infinite
	fronthaul rate and no quantization (infinite precision).

	As illustrated in Fig.~\ref{fig:u4random}, an \gls{evm} availability of $100\%$ is
	achieved with the 1-bit radio-over-fiber fronthaul architecture considered in
	this paper for the case of $B=64$ \glspl{ap}, provided that the fronthaul rate
	is at least $\SI{1382.4}{Gbit/s}$.
	The distributed massive \gls{mimo} architecture with $B=64$ \glspl{ap} significantly
	outperforms the co-located massive \gls{mimo} architecture, whose \gls{evm}
	availability, which is $30\%$, is
	derived under the optimistic assumption of no
	quantization and infinite fronthaul rate.
	The distributed \gls{mimo} architecture with $B=16$ \glspl{ap} also outperforms
	the co-located architecture when the total fronthaul rate is at least
$\qty{1382.4}{Gbit/s}$.
	However, this architecture is not able to provide an \gls{evm} availability
	close to $100\%$ for the total fronthaul rates considered in the figure.
	Finally, for the distributed \gls{mimo} architecture with $B=4$ \glspl{ap}, the
	required \gls{evm} target is not achieved for any of the random user
	placements
	considered in the simulation.
	\subsection{Imperfect Channel Knowledge}\label{sec:imperfect-csi}
	In this section, we investigate the impact of imperfect channel knowledge
	on the \gls{evm}.
	We start by noting that the expressions for the \gls{evm} with \gls{mr} and
	\gls{zf} combiners can be easily adapted to the case of imperfect channel
	knowledge.
	Specifically, let $\widetilde{\matH}_{k}\in
\complexset^{B\times U}$ be the estimate of $\widehat{\matH}_{k}$ available at the receiver.
	Furthermore, let
	\begin{equation}
		\widetilde{\matR}_{\vecq}[0] = \frac{E\sub{d}}{2}\matI_{B} + \frac{1}{N} \Re\lefto\{ \sum_{k\in \setS}
		E\sub{s}\widetilde{\matH}_{k}\herm{\widetilde{\matH}}_{k}+N\sub{0} \matI_{B}  \right\}
	\end{equation}
	and set
	\begin{equation}
		\widetilde{\matG} = \sqrt{\frac{2}{\pi}} \diag \lefto(\widetilde{\matR}_{\vecq}[0]\right)^{-1/2}.
	\end{equation}
	Finally, let
	\begin{equation}
		\widetilde{\matF}_{k} =   \left(\herm{\widetilde{\matH}}_k\widetilde{\matH}_k\right)^{-1}
		\herm{\widetilde{\matH}_k}
	\end{equation}
	and
	\begin{equation}
		\matQ_{k} = \widetilde{\matF}_{k} \widetilde{\matG}^{-1} \matG \widehat{\matH}_{k} - \matI.
	\end{equation}
	Then, for the case of Bussgang \gls{zf} combiner, we have that
	\begin{IEEEeqnarray}{rCl}
		\eta^{2} &=& \frac{1}{E\sub{s}US}  \sum_{k\in \setS} \biggl( E\sub{s}
		\Ex{}{\tr\lefto(\matQ_{k}\herm{\matQ_{k}}\right)}\notag\\
		&&+
		(N_{0}+E\sub{d})\Ex{}{\tr\lefto(\widetilde{\matF}_{k}\widetilde{\matG}^{-2}\matG^{2}\herm{\widetilde{\matF}}_{k}
			\right)} \notag\\
		&&+ \Ex{}{\tr\mathopen{}\left( \widetilde{\matF}_k\widetilde{\matG}^{-1}
			\matC_{\hat{\vece}_k}\widetilde{\matG}^{-1}\herm{\widetilde{\matF}_k} \right)}
		\biggr).\IEEEeqnarraynumspace\label{eq:evm-chest}
	\end{IEEEeqnarray}
	A similar expression can be obtained to the Bussgang \gls{mr} combiner.
	%
	One could also derive an \gls{lmmse} combiner that takes into account of the
	channel estimation errors.
	However, such a combiner admits a closed-form expression only under certain
	assumptions on $\widetilde{\matH}_{k}$ (e.g., when it is obtained via
	minimum mean square error estimation, and when $\widetilde{\matH}_{k}$
	and $\widehat{\matH}_{k}$ are jointly Gaussian---see~\cite[eqs.~(26) and (28)]{lapidoth02-05a}), which typically do not
	hold for the case of $1$-bit quantization.
	Hence, we will not pursue this direction.

	%
	In the remainder of the section, we return to the case $U=1$, $B=4$,
	assume that $R\sub{fh}=\qty{43.2}{Gbit/s}$ and focus on the Bussgang \gls{lmmse}
	channel estimator proposed in~\cite{li17-08a}.
	\begin{figure}[t]
		\centering
		\includegraphics[width=0.4\textwidth]{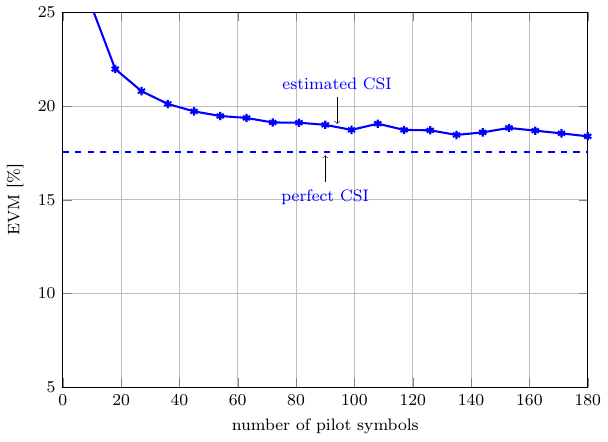}
		\caption{\gls{evm} as a function of the number of pilot symbols: $U=1$, $B=4$,
			$R\sub{fh}=\qty{43.2}{Gbit/s}$, \gls{zf} combiner, and Bussgang \gls{lmmse}
			channel estimator.
			The dashed line denotes the \gls{evm} achievable with perfect \gls{csi}.}
		\label{fig:chest}
	\end{figure}
	In Fig.~\ref{fig:chest}, we illustrate the \gls{evm} achievable in this scenario, computed
	using~\eqref{eq:evm-chest},
	as a function of the number of pilot symbols.
	We allow the power of the dither signal in the channel-estimation phase
	and in the data-transmission phase to be different, and optimize both values.
	For comparison, we also depict the \gls{evm} achievable when the channel is
	perfectly known to the \gls{cpu}.
	Not surprisingly, as the number of pilot symbols increase, the \gls{evm}
	decreases and approaches, although slowly, the one corresponding to perfect
	channel knowledge.
	Perhaps more interesting, the optimal $E\sub{d}/N_{0}$ for the
	pilot-transmission phase is
	much larger than that for the data-transmission phase (which is
$\qty{-3}{dB}$), and increases monotonically with the number of pilots.
	Specifically, it equals $\qty{5}{dB}$ when $9$ pilots are transmitted, and
	reaches $\qty{10}{dB}$ for the case of $180$ pilots.
	This is in agreement with the observation, reported in~\cite{ettefagh19-11a},
	that the  normalized mean-squared error achieved by the Bussgang \gls{lmmse}
	channel estimator decreases as a function of the number of pilot symbols only
	if the SNR before quantization is decreased simultaneously.

	\section{Conclusions}
	We have provided a characterization of the \gls{evm} achievable in the uplink of a
	distributed massive \gls{mimo} architecture with 1-bit radio-over-fiber
	fronthaul.
	Our analysis allows one to characterize the impact of a fronthaul constraint on performance and
	to shed light on the optimal design of the dither signal and on the role of spatial and temporal oversampling in such an
	architecture.
	In particular, our numerical results illustrate that, when linear combiners are
	used at the \gls{cpu}, and the \glspl{ue} transmitted
	power is assumed to be inversely proportional to the number of \glspl{ap},
	spatial oversampling is preferable to temporal oversampling for sufficiently
	large fronthaul rate.
	This is because spatial oversampling reduces multiuser interference, and, for
	the path-loss model considered in our simulations, yields a lower
	\gls{evm}, even if only a single user is active, in the limit of large fronthaul
	rates.
	However, when the fronthaul rate is small, a single user is active, and \gls{mr}
	combiner is used, temporal oversampling may be preferable to spatial
	oversampling, since it results in less correlated quantization noise.

	Our analysis reveals that the gains of distributed \gls{mimo} deployments over
	co-located deployments can be unleashed with the 1-bit radio-over-fiber
	fronthaul architecture considered in the paper, despite the nonlinearity
	introduced by the $1$-bit quantization operation.
	In future works, we will generalize our analysis to the case in which the \glspl{ap} are
	equipped with multiple antennas, and the case in which the signal exchanged between the \glspl{ap} and the \gls{cpu} has
	more than two levels.
	This requires an accurate modeling of the distortion occurring over the
	fiber-optical fronthaul.

	\appendices
	\section{Proof of Lemma~\ref{lem:ruv}}\label{app:lem}
	To prove~\eqref{eq:ruu0}, we note that, for $b=b'$ and $m=0$, the random variable $s_{bb'}[m]$
	in~\eqref{eq:buv} reduces to
	\begin{IEEEeqnarray}{rCl}
		s_{bb'}[0]&=& \frac{\frac{c[0]}{N }\bigl( E\sub{s}\vecnorm{\vech_{b}}^{2}+ N_{0}\bigr) + \frac{E\sub{d}  }{2}
		}{p_{b}^{2} } \\
		&=&\frac{\frac{S}{N }\bigl( E\sub{s}\vecnorm{\vech_{b}}^{2}+ N_{0}\bigr) + \frac{E\sub{d}  }{2}
		}{p_{b}^{2} }
		= 1.
	\end{IEEEeqnarray}
	Here, the second equality follows because $c[0]=S$
	(see~\eqref{eq:cm}), and the
	third equality follows from the definition of $p_{b}$
	in~\eqref{eq:pu}.
	Next, we establish that $s_{bb'}[m]\leq s_{bb}[0] $.
	The desired result then follows from~\eqref{eq:ruv-def}, from the
	monotonicity of the function $f(x)=\arcsin(x)-x$, and because $f(1)=\pi/2-1$.
	Note that, when $b\neq b'$ or $m\neq 0$,
	\begin{IEEEeqnarray}{rCl}
		s_{bb'}[m] &\leq& \frac{\frac{S}{N}\abs{E\sub{s} \tp{\vech_{b}} \conj{\vech_{b'}} +N_{0}}}{p_{b} p_{b'}} \\
		&\leq & \frac{\frac{S}{N}\bigl(E\sub{s} \vecnorm{\vech_{b}}\vecnorm{\vech_{b'}} +N_{0}\bigr) }{p_{b}p_{b'}}
		\leq 1.\label{eq:triangle-cs-for-buv}
	\end{IEEEeqnarray}
	Here, in the first inequality we used that, for all $x\in\reals$, $\Re\{x\}\leq
\abs{x}$; the second inequality is a
	consequence of triangle and Cauchy-Schwarz inequalities; finally, in the last inequalities we have lower-bounded the
	denominator and, hence, upper-bounded the ratio, by neglecting the term $E\sub{d}/2$ in $p_{b}$ and $p_{b'}$; we also
	used that for all nonnegative $a$, $b$, $c$,
	\begin{equation}
		\frac{ab+c}{\sqrt{a^2+c}\sqrt{b^{2}+c}}\leq 1.
	\end{equation}

	To establish~\eqref{eq:ruvm}, we note that, when $b\neq b'$ or $m\neq 0$, we
	have that
	\begin{multline}
		\abs{s_{bb'}[m] } \leq \frac{\frac{S}{N}\bigl(E\sub{s} \vecnorm{\vech_{b}}\vecnorm{\vech_{b'}} +N_{0}\bigr) }{p_{b}p_{b'}}
		\overset{\text{a.s.}}{\longrightarrow} 0, \\
		N\to\infty,\quad
		u=1,\dots,B.
	\end{multline}
	The last step follows because, as $N\to\infty$, the numerator vanishes, whereas the denominator converges to
$E\sub{d}/2$.

	To establish~\eqref{eq:sum-ruvm}, we note that, as a consequence
	of~\eqref{eq:ruv-def},
	of the monotonicity of $f(x)$, and of its
	symmetry,
	\begin{IEEEeqnarray}{rCl}\label{eq:ruv-ub}
		\abs{r_{bb'}[m] } = \abs{f(s_{bb'}[m]) } = f(\abs{s_{bb'}[m]}).
	\end{IEEEeqnarray}
	Also note that for $m\geq 1$ (and also when $m=0$ but $b\neq b'$),
	\begin{equation}\label{eq:buv-ub}
		\abs{s_{bb'}[m]} \leq  \min\lefto\{1,\frac{2S\bigl(E\sub{s}
			\vecnorm{\vech_{b}}\vecnorm{\vech_{b'}}+N_{0}\bigr)}{NE\sub{d}}\right\}.
	\end{equation}
	Here, the inequality is obtained by proceeding as in~\eqref{eq:triangle-cs-for-buv} and then using that
$p_{b}\geq \sqrt{E\sub{d}/2}$.
	Hence,
	\begin{IEEEeqnarray}{rCl}
		&&\sum_{m=1}^{N-1} \abs{r_{bb'}[m]}  \notag \\
		&&\leq\sum_{m=1}^{N-1} f\lefto( \min\lefto\{1,\frac{2S\bigl(E\sub{s}
			\vecnorm{\vech_{b}}\vecnorm{\vech_{b'}}+N_{0}\bigr)}{NE\sub{d}}\right\}
		\right)\\
		&&= N  f\lefto( \min\lefto\{1,\frac{2S\bigl(E\sub{s}
			\vecnorm{\vech_{b}}\vecnorm{\vech_{b'}}+N_{0}\bigr)}{NE\sub{d}}\right\}\label{eq:sum-ruv-ub}
		\right).
	\end{IEEEeqnarray}
	The desired result follows because
	\begin{equation}
		\lim_{N\to\infty} N^{3}f\lefto(1/N\right) = 1/6
	\end{equation}
	from
	which~\eqref{eq:sum-ruvm} follows via~\eqref{eq:sum-ruv-ub}.
	\section{Proof of Theorem~\ref{thm:as-evm}}\label{app:thm}
	First note that, for the case $U=1$, the Bussgang \gls{mr} and \gls{zf} combiners
	coincide.
	Let us first assume that we can interchange limit and expectation.
	Then, specializing~\eqref{eq:symbol-estimate-zf} for the case of frequency-flat
	fading and $U=1$, and using~\eqref{eq:limit-matG} and~\eqref{eq:limit-Ce},
	we conclude that
	\begin{multline}\label{eq:intermediate-asymptotic-mr-zf-u1}
		\lim_{N\to\infty}\eta^{2} = \frac{1}{E\sub{s}} \biggl( (N_{0}+E\sub{d})
		\Ex{}{\frac{1}{\vecnorm{\vech}^{2}}} \\ + 2 \left( 1 - \frac{2}{\pi}\right)
		\frac{\pi E\sub{d}}{4}\Ex{}{\frac{1}{\Ex{}{\vecnorm{\vech}^{2}}}}\biggr),
	\end{multline}
	from which~\eqref{eq:evm-mr-as} follows after simplifications.

	To show that expectation and limit can be interchanged,
	we rewrite $\eta^{2}$ for finite $N$ as
	\begin{IEEEeqnarray}{rCl}
		\eta^{2} &=& \frac{N_{0} + E\sub{d} }{E\sub{s}}\Ex{}{\frac{1}{\vecnorm{\vech}^{2}}} +
		\frac{1}{E\sub{s} } \sum_{b=1}^{B}
		\Ex{}{\frac{2}{\vecnorm{\vech}^{4}} \abs{h_{b} }^{2} p_{b}^{2}r_{bb}[0]
		}\notag\\
		&&+
		\frac{1}{E\sub{s} } \sum_{b=1}^{B} \sum_{b'\neq b}
		\Ex{}{\frac{2}{\vecnorm{\vech}^{4}} \conj{h_{b}}h_{b'}p_{b}p_{b'}
		r_{bb'}[0]}\notag\\
		&&+
		\frac{1}{SE\sub{s} } \sum_{m=1}^{N-1} \sum_{b=1}^{B} \sum_{b'=1}^{B}
		\Ex{}{\frac{2}{\vecnorm{\vech}^{4}} \conj{h_{b}}h_{b'}p_{b}p_{b'}
		r_{bb'}[m]} \notag\\
		&&\quad \times c[m] e^{-j2\pi \frac{f_{c}}{f_{s}}m}.\label{eq:split-mr}
	\end{IEEEeqnarray}
	Next, we use the
	dominated convergence theorem~\cite[Thm.~1.34]{rudin87a} three times, to show
	that the limit $N\to\infty$ can be interchanged with the expectation in the last
	three terms comprising~\eqref{eq:split-mr}.
	Specifically, in the second term in~\eqref{eq:split-mr}, the interchange can be
	performed if there exists a random variable
$\gamma_{b}$, $b\in \{1,\dots,B\}$, not depending on $N$ and satisfying
	\begin{equation}\label{eq:expectation-1-mr}
		\Ex{}{\abs{h_{b}}^{2}\gamma^{2}_{b}  /\vecnorm{\vech}^{4}}<\infty
	\end{equation}
	and
	\begin{equation}\label{eq:inequality-part1-mr}
		p_{b}^{2} \leq \gamma^{2}_{b} \quad \text{a.s.} \quad \text{for all } b,
		\text{ and } N.
	\end{equation}
	To obtain this random variable, we set
	\begin{equation}
		\gamma_{b}= \sqrt{S (E\sub{s}\abs{h_{b}}^{2}+N_{0}) + E\sub{b} /2}
	\end{equation}
	and note that $p_{b}\leq \gamma_{b}$ for all $N$.
	Furthermore,~\eqref{eq:expectation-1-mr} follows
	from~\eqref{eq:technical-cond-mr}.

	In the third term in~\eqref{eq:split-mr}, the interchange can be performed if
	there exist random variables
$t_{bb'}$, $b,b'\in\{1,\dots, B\}$ with $b\neq b'$ not depending on $N$ and satisfying
	\begin{equation}\label{eq:expectation-part2-mr}
		\Ex{}{\conj{h}_{b} h_{b'} t_{bb'}  /\vecnorm{\vech}^{4}}<\infty
	\end{equation}
	and
	\begin{equation}\label{eq:inequality-part-2-mr}
		\abs{p_{b}p_{b'}r_{bb'}[0]}\leq t_{bb'}, \quad \text{a.s.}
	\end{equation}
	for all $N$ and all $b,b'$ with $b\neq b'$.
	To construct these random variables, we use~\eqref{eq:ruu0}
	to conclude that
	\begin{equation}\label{eq:ruv0-ub}
		\abs{r_{bb'}[0] }\leq \pi/2-1.
	\end{equation}
	It follows from~\eqref{eq:ruv0-ub} and~\eqref{eq:inequality-part1-mr} that~\eqref{eq:inequality-part-2-mr} holds with
$t_{bb'}$ given in~\eqref{eq:def-tbb}.
	Finally,~\eqref{eq:expectation-part2-mr} follows
	from~\eqref{eq:technical-cond-mr}.

	In the fourth term in~\eqref{eq:split-mr}, the interchange can be
	performed if there exist random variables
$\bar{t}_{bb'}$, $b,b'\in \{1,\dots,B\}$ not depending on $N$ and satisfying
	\begin{equation}\label{eq:expectation-part3-mr}
		\Ex{}{\frac{\conj{h}_{b} h_{b'}  }{\vecnorm{\vech}^{4}}\bar{t}_{bb'} }<\infty
	\end{equation}
	and
	\begin{equation}\label{eq:inequality-part3-mr}
		\abs{\sum_{m=1}^{N-1} p_{b}p_{b'}r_{bb'}[m]c[m]e^{-j2\pi \frac{f_{c} }{f_{s}
		}m}} \leq \bar{t}_{bb'}  \quad \text{a.s.} \, \text{for all } b,b',N.
	\end{equation}
	Proceeding as in the previous parts, we observe that
	\begin{IEEEeqnarray}{rCl}
		&&\abs{\sum_{m=1}^{N-1} p_{b}p_{b'}r_{bb'}[m]c[m]e^{-j2\pi \frac{f_{c} }{f_{s}
		}m}}\notag\\
		&\leq&
		\gamma_{b}\gamma_{b'} S\sum_{m=1}^{N-1} \abs{r_{bb'}[m]}
		\label{eq:inequality-part-3-mr-derivation-p1}\\
		&\leq& \gamma_{b}\gamma_{b'} SN f\lefto(\min\left\{1,\frac{2SE\sub{s}
			\abs{h_{b}}\abs{h_{b'}}+N_{0}}{NE\sub{d}}\right\}\right)\\
		&\leq&  \left( \frac{\pi}{2}-1 \right)S \gamma_{b}\gamma_{b'}
		\left(\frac{2SE_{s}\abs{h_{b}} \abs{h_{b'}} +N_{0}}{E\sub{d}}\right). \label{eq:inequality-part-3-mr-derivation}
	\end{IEEEeqnarray}
	Here, in the first inequality we used~\eqref{eq:inequality-part1-mr} and the
	triangle inequality; the second inequality follows from~\eqref{eq:sum-ruv-ub};
	in the third inequality, we used that the function $f(x)/x$ is monotonically
	increasing in $x$ and that $f(x)\leq (\pi/2-1)x$.
	It then follows from~\eqref{eq:inequality-part-3-mr-derivation}
	that~\eqref{eq:inequality-part3-mr}
	holds with  $\bar{t}_{bb'}$ given in~\eqref{eq:def-tbb'}.
	Finally,~\eqref{eq:expectation-part3-mr} follows
	from~\eqref{eq:technical-cond-mr-2}.

	We now prove~\eqref{eq:evm-mmse-as}.
	If limit and expectation can be interchanged, then~\eqref{eq:evm-mmse-as}
	follows directly from~\eqref{eq:evm-mmse},~\eqref{eq:limit-matG},
	and~\eqref{eq:limit-Ce}.
	To prove that the expectation and the limit can indeed be interchanged, we use
	again the dominated convergence theorem.
	We start by noting that, for the frequency-flat case and when $U=1$, the
	\gls{evm} expression~\eqref{eq:evm-mmse} simplifies to
	\begin{equation}\label{eq:evm-u1-l1-mmse}
		\eta^{2} = \frac{1}{S} \sum_{k\in\setS} \Ex{}{\frac{1}{1+z_{k} }},
	\end{equation}
	where
	\begin{equation}
		z_{k} =  \herm{\vech} \left(\frac{N_0+E\sub{d}}{E\sub{s}} \matI_B +
		\frac{1}{E\sub{s}}\matG^{-1}\matC_{\hat{\vece}_k}\matG^{-1}
		\right)^{-1}\vech,
		\label{eq:zk-single-user}
	\end{equation}
	with the entry on the $b$th row and $b'$th column of the matrix $\matG^{-1}\matC_{\hat{\vece}_k}\matG^{-1}
$ given by
	\begin{multline}\label{eq:mmse-elements-matrix}
		\Bigl[ \matG^{-1}\matC_{\hat{\vece}_k}\matG^{-1}  \Bigr]_{b,b'} =\\
		2\sum_{m=0}^{N-1} p_{b}p_{b'} r_{bb'}[m] e^{-j2\pi (k/N+f\sub{c}/f\sub{s})m
			},\\  b=1,\dots,B, b'=1,\dots,B.
	\end{multline}
	We shall show that there exists a random variable $\bar{\gamma}$
	that does not depend on $N$ and satisfies
	\begin{equation}\label{eq:expectation-mmse-evm}
		\Ex{}{1/(1+\bar{\gamma}) }<\infty
	\end{equation}
	and
	\begin{equation}\label{eq:inequality-mmse}
		\frac{1}{1+z_{k} } \leq \frac{1}{1+\bar{\gamma}} \quad \text{a.s.} \quad \text{for all }
		N\text{ and }k\in\setS.
	\end{equation}
	Let
	\begin{equation}
		\matP_{k}=\frac{N_0+E\sub{d}}{E\sub{s}} \matI_B +
		\frac{1}{E\sub{s}}\matG^{-1}\matC_{\hat{\vece}_k}\matG^{-1}.
	\end{equation}
	By construction, $\matP_{k} $ is Hermitian and positive definite.
	As a consequence, $\matP_{k}^{-1}$ is also Hermitian and positive definite.
	Let $\lambda\sub{min}(\matP_{k}^{-1})$ denote the smallest eigenvalue of
$\matP_{k} ^{-1}$, and $\lambda\sub{max}(\matP_{k} )$ denote the largest eigenvalue of
$\matP_{k}$.
	Note that, almost surely,
	\begin{IEEEeqnarray}{rCl}
		\herm{\vech}\matP_{k} ^{-1}\vech &\geq&
		\lambda\sub{min}(\matP_{k} ^{-1})\vecnorm{\vech}^{2} =
		\frac{1}{\lambda\sub{max}(\matP_{k} ) } \vecnorm{\vech}^{2}\\ &\geq&
		\frac{1}{\tr(\matP_{k})}\vecnorm{\vech}^{2}.
	\end{IEEEeqnarray}
	Here, the first inequality follows from the Rayleigh-Ritz
	Theorem~\cite[Thm.~4.2.2]{horn85a} and the second inequality follows because
	the trace of $\matP_{k}$ is equal to the sum of the eigenvalues of $\matP_{k}$.
	Finally, note that, as a consequence of~\eqref{eq:mmse-elements-matrix},
	\eqref{eq:inequality-part1-mr}, \eqref{eq:inequality-part-2-mr},
	and proceeding similarly to~\eqref{eq:inequality-part-3-mr-derivation-p1}--\eqref{eq:inequality-part-3-mr-derivation},
	\begin{IEEEeqnarray}{rCl}
		\tr(\matP_{k}) &=& \frac{N_{0} +E\sub{d} }{E\sub{s} }B \notag \\ &&+ \frac{2}{E\sub{s} }
		\sum_{b=1}^{B}\sum_{b'=1}^{B}   \sum_{m=0}^{N-1} p_{b}p_{b'} r_{bb'}[m] e^{-j2\pi (k/N+f\sub{c}/f\sub{s})m
			}\notag\\
		&\leq& \frac{N_{0} +E\sub{d} }{E\sub{s} }B     + \frac{2}{E\sub{s}
		}\sum_{b=1}^{B}\gamma_{b}^{2}
		\left( \frac{\pi}{2}-1 \right)\notag \\
		&&+ \frac{2}{E\sub{s} }
		\sum_{b=1}^{B}\sum_{b'=1}^{B} t_{bb'} +
		\frac{2}{E\sub{s}S}\sum_{b=1}^{b}\sum_{b'=1}^{B} \bar{t}_{bb'}  .
	\end{IEEEeqnarray}
	As a consequence, we conclude that~\eqref{eq:inequality-mmse} holds with
$\bar{\gamma}$ given in~\eqref{eq:bar-gamma}.
Finally, we note that~\eqref{eq:expectation-mmse-evm} coincides with assumption~\eqref{eq:technical-cond-mmse}.
\bibliographystyle{IEEEtran}
\bibliography{extracted}
\end{document}